\documentclass[floatfix,
reprint,
superscriptaddress,
amsmath,amssymb,
aps,
prl,
]{revtex4-2}

\usepackage{graphicx}
\usepackage[dvipsnames]{xcolor}
\usepackage{pdfpages}
\usepackage{amssymb}
\usepackage{amsmath}
\usepackage{amsbsy}
\usepackage{color}
\usepackage[normalem]{ulem}
\usepackage{epstopdf}
\usepackage{bm}
\usepackage{xurl}
\usepackage{textcomp}

\usepackage{float}

\usepackage{filecontents}
\makeatletter\@input{z2.tex}\makeatother

\graphicspath{{figures/}}

\usepackage{xr}
\makeatletter
\AtBeginDocument{\let\LS@rot\@undefined}
\newcommand*{\addFileDependency}[1]{
  \typeout{(#1)}
  \@addtofilelist{#1}
  \IfFileExists{#1}{}{\typeout{No file #1.}}
}
\makeatother
\newcommand*{\myexternaldocument}[1]{%
    \externaldocument[SM-]{#1}%
    \addFileDependency{#1.tex}%
    \addFileDependency{#1.aux}%
}
\myexternaldocument{SI}

\begin{document}
\title{Designing DNA nanostar hydrogels with programmable degradation and antibody release}

\author{Giorgia Palombo}
\affiliation{School of Physics and Astronomy, University of Edinburgh, Peter Guthrie Tait Road, Edinburgh, EH9 3FD, UK}

\author{Christine A. Merrick}
\affiliation{Centre for Engineering Biology, School of Biological Sciences, University of Edinburgh, Edinburgh, UK}

\author{Jennifer Harnett}
\affiliation{School of Physics and Astronomy, University of Edinburgh, Peter Guthrie Tait Road, Edinburgh, EH9 3FD, UK}

\author{Susan Rosser}
\affiliation{Centre for Engineering Biology, School of Biological Sciences, University of Edinburgh, Edinburgh, UK}

\author{Davide Michieletto}
\thanks{corresponding author, davide.michieletto@ed.ac.uk}
\affiliation{School of Physics and Astronomy, University of Edinburgh, Peter Guthrie Tait Road, Edinburgh, EH9 3FD, UK}
\affiliation{MRC Human Genetics Unit, Institute of Genetics and Cancer, University of Edinburgh, UK}
\affiliation{International Institute for Sustainability with Knotted Chiral Meta Matter (WPI-SKCM\textsuperscript{2}), Hiroshima University, Higashi-Hiroshima, Hiroshima 739-8526, Japan}

\author{Yair Augusto Guti\'{e}rrez Fosado}
\thanks{corresponding author, yair.fosado@ed.ac.uk}
\affiliation{School of Physics and Astronomy, University of Edinburgh, Peter Guthrie Tait Road, Edinburgh, EH9 3FD, UK}

\begin{abstract}
\textbf{DNA nanostar (DNAns) hydrogels are promising materials for \textit{in vivo} applications, including tissue regeneration and drug and antibody delivery. However, a systematic and quantitative understanding of the design principles controlling their degradation is lacking. Here, we investigate hydrogels made of three-armed DNAns with varying flexible joints, arm lengths, and mesh sizes and use restriction enzymes (RE) to cut the DNAns structures while monitoring the gel's degradation. We discover that (i) removing flexible joints, (ii) increasing arm length, or (iii) relocating the RE site along a DNA linker markedly accelerates hydrogel degradation. In contrast, non-specific endonucleases, e.g. DNaseI, quickly degrade DNAns hydrogels regardless of design. Importantly, the release of antibodies from DNAns hydrogels can be modulated by the action of sequence-specific enzymes, confirming that programmable degradation can be leveraged for responsive drug-delivery systems. These findings provide a better understanding of the design principles for DNAns-based scaffolds with tunable degradation, cargo release, and responsive rheology.
}
\end{abstract}

\maketitle

\begin{figure*}[t!]
\centering
\includegraphics[width=0.92\linewidth]{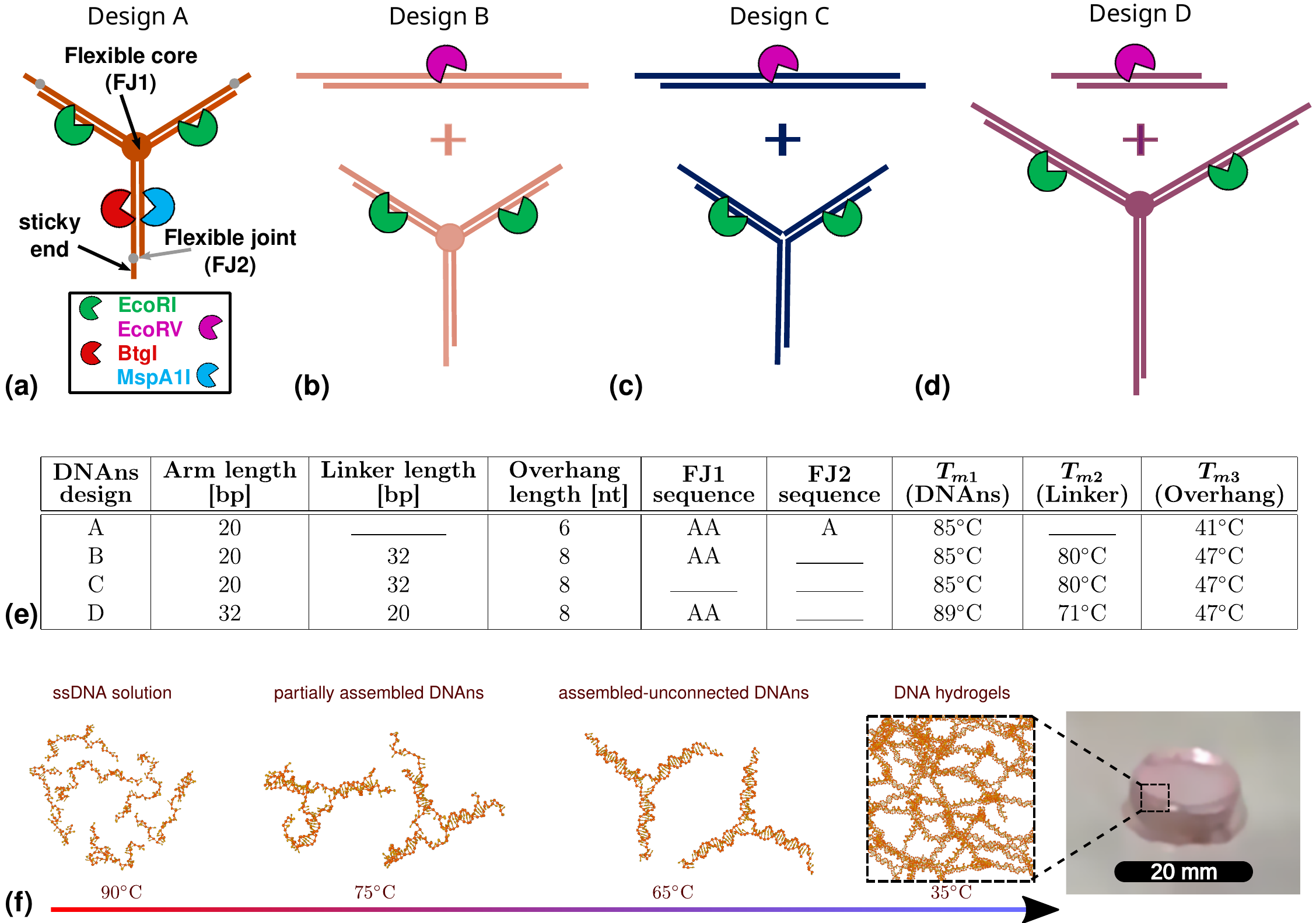} 
\caption{\textbf{DNA nanostar  designs}. \textbf{(a)-(d)} Schematic representations of the four DNAns designs used in this study. Each design self-assembles into a three-armed DNA nanostar containing recognition sequences for the restriction enzymes EcoRI-HF (two arms, labeled in green), BtgI, and MspA1I (remaining arm, labeled in red and cyan, respectively). For designs B-D an EcoRV-HF site (labelled in magenta) is located on the linker. \textbf{(e)} Summary table of structural parameters for each design. Columns two to four list arm, linker, and sticky-end lengths (in base-pairs (bp) and nucleotides (nt)). Columns five and six indicate the number of unpaired adenines (A) forming flexible joints at the core (FJ1) and before the sticky ends (FJ2). The final three columns report the melting temperatures determined with NUPACK~\cite{nupack} at [NaCl]=150mM, [Mg$^{+2}$]=10mM and [DNA]=250$\mu$M. \textbf{(f)} Schematic of the annealing protocol used for DNAns and hydrogel assembly (shown for design A, without linker).
}
\label{fig:model}
\end{figure*}

Deoxyribonucleic acid (DNA) nanotechnology has rapidly progressed from fundamental research to applied technology~\cite{Seeman2017}. Its unmatched molecular specificity and programmability make DNA an ideal material for constructing self-assembling, responsive, and highly tunable nanostructures~\cite{Seeman2017,Rothemund2006}. Among these, DNA nanostars (DNAns) have emerged as versatile building blocks, serving both as model systems for probing the properties of molecular liquids~\cite{Palombo2025,Biffi2013,Biffi2015,Conrad2019,Brady2018AmphiphilicDNA} and as functional materials~\cite{Stoev2025} for tissue regeneration~\cite{Yuan2021}, biosensing~\cite{English2019}, synthetic cells~\cite{Malouf2023SculptingDNA,DiMichele2024JACS,DiMichele2025}, and drug delivery~\cite{Veneziano2021,Fabrini2022QuadStars}. DNAns are formed from short single-stranded DNA (ssDNA) oligomers designed to hybridize into multi-armed, double-helical motifs emanating from a central junction~\cite{Um2006} (Fig.~\ref{fig:model}). The arms terminate in short self-complementary ssDNA overhangs, or ``sticky ends'', which mediate the assembly of adjacent DNAns into higher-order networks. By programming arm number (or valence), arm length, junction flexibility, and overhang length, DNAns can be precisely tuned to control the mechanical properties, connectivity, and responsiveness of the resulting material~\cite{Xing2018,Nguyen2017,Palombo2025,Conrad2019,Fernandez-Castanon2018,Fosado2023}. At high concentrations, DNAns percolate into soft, hydrophilic networks known as DNA hydrogels~\cite{Um2006,Xing2018,Biffi2013,Nguyen2017,Locatelli2017,Conrad2019,Palombo2025}, whose programmable structure offers unique opportunities for controlled biomolecule encapsulation, triggered release, and responsive degradation, key features for the next-generation of biomedical applications.

The mechanical properties of DNAns hydrogels can be tuned \textit{via} chemical and physical methods~\cite{Veneziano2021}, as well as by modulating the specific structure and connectivity of the DNAns building blocks~\cite{Fernandez-Castanon2018,Xing2018,Bomboi2016,Stoev2020}. In addition, nature provides a diverse toolkit of enzymes capable of manipulating DNA with high precision and modifying the ensuing material properties in dense and entangled solutions of long DNA molecules~\cite{Michieletto2022natcomm,Panoukidou2022,Conforto2024,Fosado2022}. However, the precise rheological control of DNAns hydrogels through DNA-altering enzymes has so far been largely unexplored despite being a promising route to programmable degradation \textit{in vivo}. 

To address this gap, here we investigate the enzymatic degradation of DNAns hydrogels using restriction enzymes (REs) and systematically quantify the degradation of, and antibody release from, DNAns hydrogels. Notably, while the enzymatic degradation of DNAns liquid droplets has been comprehensively examined in recent studies~\cite{Saleh2020PNAS,Saleh2023}, the degradation of the ``gel'' regime is unexplored.

By combining confocal microscopy, time-resolved microrheology, and gel electrophoresis, we investigated the enzymatic degradation of DNAns hydrogels made with different DNAns designs and across different phases. 

One of the key findings of our work is that, under the experimental conditions tested, non-specific endonucleases such as DNaseI readily induce network cleavage, whereas site-specific REs show limited impact on gel integrity. By systematically testing different structural designs, we discovered that dense connectivity and restricted internal dynamics of DNAns networks sterically hinder enzyme access to recognition sites, thereby providing intrinsic protection against sequence-specific cleavage. Importantly, this steric hindrance can be modulated by nanostar design, enabling programmable control over enzymatic accessibility and degradation rates.

Finally, leveraging this programmable stability, we demonstrate loading and enzymatic release from hydrogels of humanized monoclonal IgG, a model antibody currently used in clinical trials as a treatment for cancer. These findings establish a design-based strategy for engineering DNA materials with programmable degradation and release, advancing the development stimuli-responsive scaffolds for applications in tissue regeneration, 3D cell culture, and drug delivery systems with continuous time-varying rheology.

\begin{figure*}[t!]
\centering
\includegraphics[width=0.85\linewidth]
{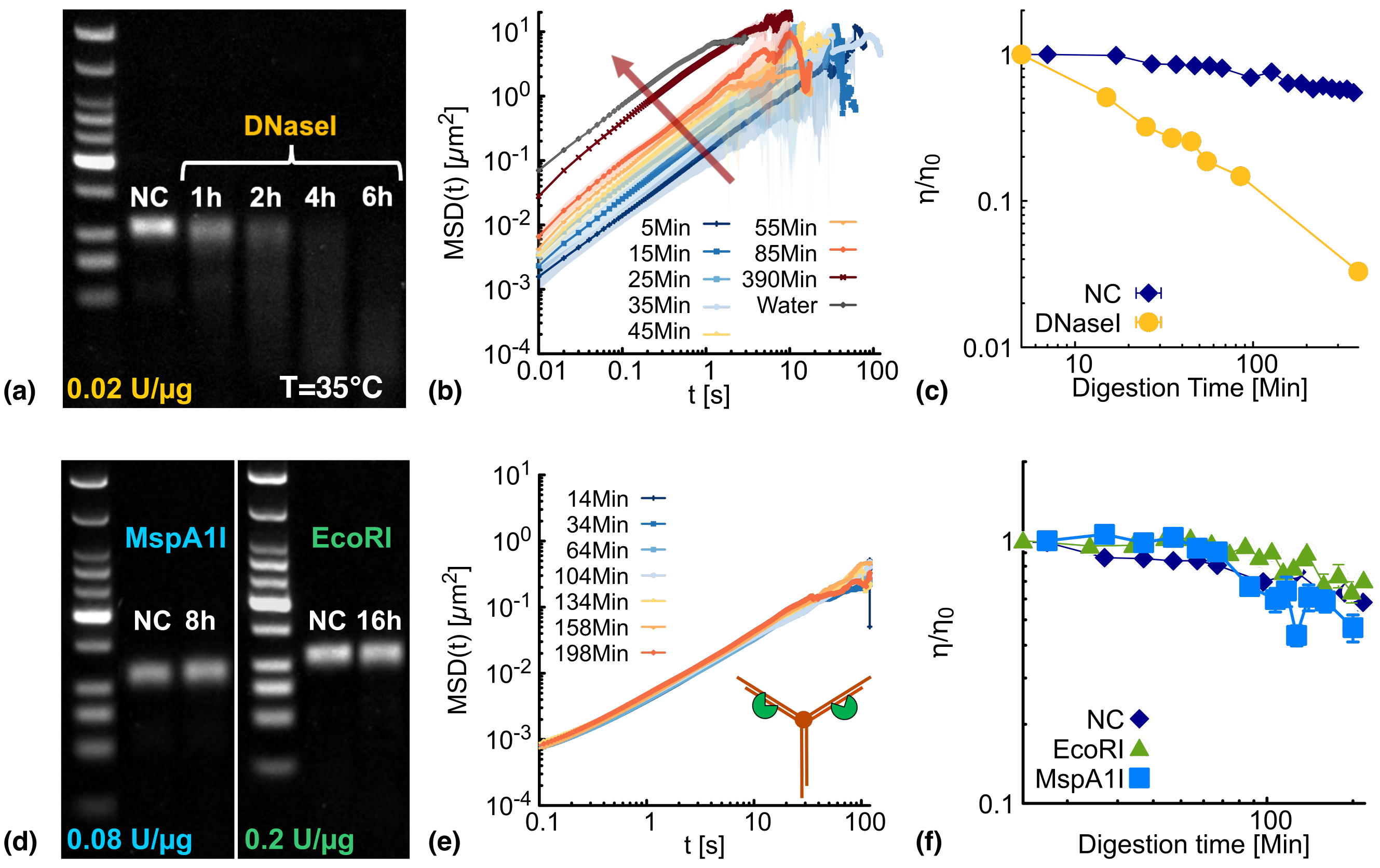}
\caption{\textbf{DNA hydrogels formed by design A nanostars display RE resistance.} \textbf{(a)} Agarose gel electrophoresis (native conditions) of DNA fragments from DNAns hydrogels (250$\mu$M, corresponding to $\sim120\mu$g/$\mu$L) treated with DNaseI for different incubation times (1, 2, 4 and 6 hours) at 35$^\circ$C. ``NC'' is negative control (no addition of RE). Ladder is NEB low molecular weight. \textbf{(b)} Time-resolved microrheology shows that the gel progressively loses viscoelasticity. \textbf{(c)} Normalized viscosity decreases monotonically with time. \textbf{(d)} Agarose gel electrophoresis (native conditions) of DNAns hydrogels (at 250$\mu$M) treated with the sequence-specific restriction enzymes MspA1I (8h incubation) and EcoRI (16h incubation) shows no detectable degradation. \textbf{(e)} Microrheology confirms no significant change in material properties with respect to control. \textbf{(f)} Normalized viscosity as a function of time display no significant loss of integrity. In panels (c) and (f) the viscosity at digestion time $t$ is normalized by the initial value ($\eta_0$), measured at the earliest available time point.}
\label{fig:MSD_RE}
\end{figure*}

\section{Results and discussion}
To systematically explore how nanostar structural features influence enzymatic degradation, we designed four DNAns variants, labeled A–D (see Fig.~\ref{fig:model}). In design A, the sticky ends are self-complementary, allowing direct hybridization between adjacent nanostars. In contrast, designs B–D feature non-self-complementary sticky ends that hybridize only with the overhangs of double-stranded DNA linkers, which act as bridges connecting neighboring nanostars. This architectural difference enables control over network connectivity and pore size, while variations in core and joint flexibility determine the molecular dynamics and accessibility of the restriction sites. Detailed sequences,  annealing protocols for each design and sample preparations are provided in the Methods section and Supporting Information (sections~\ref{SM-sec:designSequences}, \ref{SM-sec:DNAnsassembly} and ~\ref{SM-sec:samplePrep}).

In the following sections, we study the degradation properties of DNAns gels through a combination of gel electrophoresis and time-resolved microrheology. This dual approach allows us to visually identify digested DNA fragments and correlate the digestion with the rheological properties of the gel. 

\subsection{Linker-free DNAns hydrogels are resistant against site-specific endonucleases}
We first examined the enzymatic digestion of hydrogels formed by design A DNAns upon treatment with the non-specific and highly processive DNaseI (with an enzyme(U)-to-DNA($\mu$g) ratio of 0.02U/$\mu$g, i.e. 2U of enzyme in 120 $\mu$g of DNAns). As shown in Fig.~\ref{fig:MSD_RE}(a), the negative control (no enzyme) displays a single band, corresponding to intact design A nanostars (running around 100 bp). Upon addition of DNaseI, the bands intensity gradually decreases and eventually disappears after approximately 4 hours, indicating complete digestion of the DNAns (see SI, section~\ref{SM-sec:dnase1DiffDesigns} for DNaseI digestion of designs B-D). This observation is confirmed by microrheology measurements: Fig.~\ref{fig:MSD_RE}(b) reports the mean-squared displacement (MSD) of $a = 200$ nm tracer beads, $\textrm{MSD}(t, t_d) = \langle [\bm{r}(t + t_0; t_d) - \bm{r}(t_0; t_d)]^2 \rangle$, where the average is performed over tracers, positions in the sample and initial times $t_0$. The ``digestion time'' $t_{d} \gg t_0$ denotes the elapsed time since enzyme addition, while the lagtime $t$ denotes the time elapsed between two positions. Typically, we record 2 minutes videos every 10 minutes over several hours, allowing us to extract sequential MSD curves (MSD(t, 5 min), MSD(t, 15 min), ...) and monitor the temporal evolution of sample viscosity. As digestion proceeds, the MSD curves shift upward, reflecting a progressive decrease in viscosity, i.e. the beads diffuse faster in the hydrogel. The corresponding viscosity, $\eta (t_d) = k_B T/(3\pi D_p (t_d) a)$, where $D_p (t_d) = \lim_{t \to \infty} MSD(t; t_d)/2t$, is plotted as a function of digestion time in Fig.~\ref{fig:MSD_RE}(c). In the negative control we observe a mild reduction in viscosity (likely due to the presence of glycerol in the enzyme buffer, see SI, section~\ref{SM-sec:glyceroleffect}), on the contrary the DNaseI-treated sample displays a $\sim$ 33-fold reduction in viscosity within six hours.

We next performed analogous experiments using sequence-specific REs to test whether they could degrade design A hydrogel. We used MspA1I (0.08 U/$\mu$g, 10 U enzyme in 120 $\mu$g DNA) and EcoRI (0.2 U/$\mu$g, 20 U enzyme in 120 $\mu$g DNA). For comparison, in the phase-separation regime, DNAns droplets are completely digested by EcoRI at a much higher stoichiometry (20 U/$\mu$g, 20 U enzyme and 1 $\mu$g DNA) within approximately 800 s (see SI, Fig.~\ref{SM-fig:Confocal_NC_enzymes}). Achieving comparable enzyme-to-DNA ratios in bulk hydrogels would require enzyme quantities that are impossible to attain at the much higher DNA concentrations used here. Even when using the highest feasible enzyme concentrations, a naive linear extrapolation predicts gel degradation on a timescale of $\sim$22 hours, roughly 100-fold longer than observed for the droplets. However, no measurable digestion was detected over these timescales. Surprisingly, both gel electrophoresis (Fig.~\ref{fig:MSD_RE}(d)) and microrheology (Fig.~\ref{fig:MSD_RE}(e,f)) show that the hydrogel remains intact over time. This lack of degradation occurred despite the use of enzyme-to-DNA mass ratios four times higher for MspA1I and ten times higher for EcoRI compared to the DNaseI experiments, in which complete digestion was observed. Indeed, the behaviour of samples treated with sequence-specific enzymes is indistinguishable from that of the negative controls (Fig.~\ref{fig:MSD_RE}(f)).

We note that enzyme activity units are defined under assay-specific conditions (see SI section~\ref{SM-sec:samplePrep}) and do not directly reflect cleavage efficiency across different DNA substrates. Therefore, all comparisons in this study are based on the observed activity under our experimental conditions. Within this framework, DNaseI is nonetheless intrinsically several orders of magnitude more processive than the other REs examined here.

The resistance of DNAns hydrogels to site-specific restriction enzyme degradation was unexpected, especially given previous observations of enzymatic degradation in DNAns droplets~\cite{Saleh2020PNAS, Saleh2023} and qualitative visual effects on the stability of DNAns gels~\cite{Xing2011}. We therefore sought to understand the origin of this resistance. One possible explanation is that steric hindrance within the hydrogel limits the ability of restriction enzyme dimers ($\sim$8 nm in size~\cite{Kim1990}) to access and scan their target recognition sites~\cite{Pingoud2001}, even though the gel’s pore size ($\simeq20$ nm~\cite{Palombo2025}, comparable to the contour length distance between the cores of two hybridized DNAns~\cite{Fosado2023}) exceeds the enzyme's dimension. In contrast, DNaseI, a smaller and non-specific endonuclease ($\sim$4 nm in size~\cite{Dietrich1994}), can arguably bind and cleave DNA regions efficiently. To test this hypothesis, we first examine whether increasing the gel’s pore size would facilitate degradation. 

\subsection{Pore size does not affect the degradation efficiency of DNAns hydrogels}

\begin{figure}[!ht]
\centering
\includegraphics[width=1.0\columnwidth]{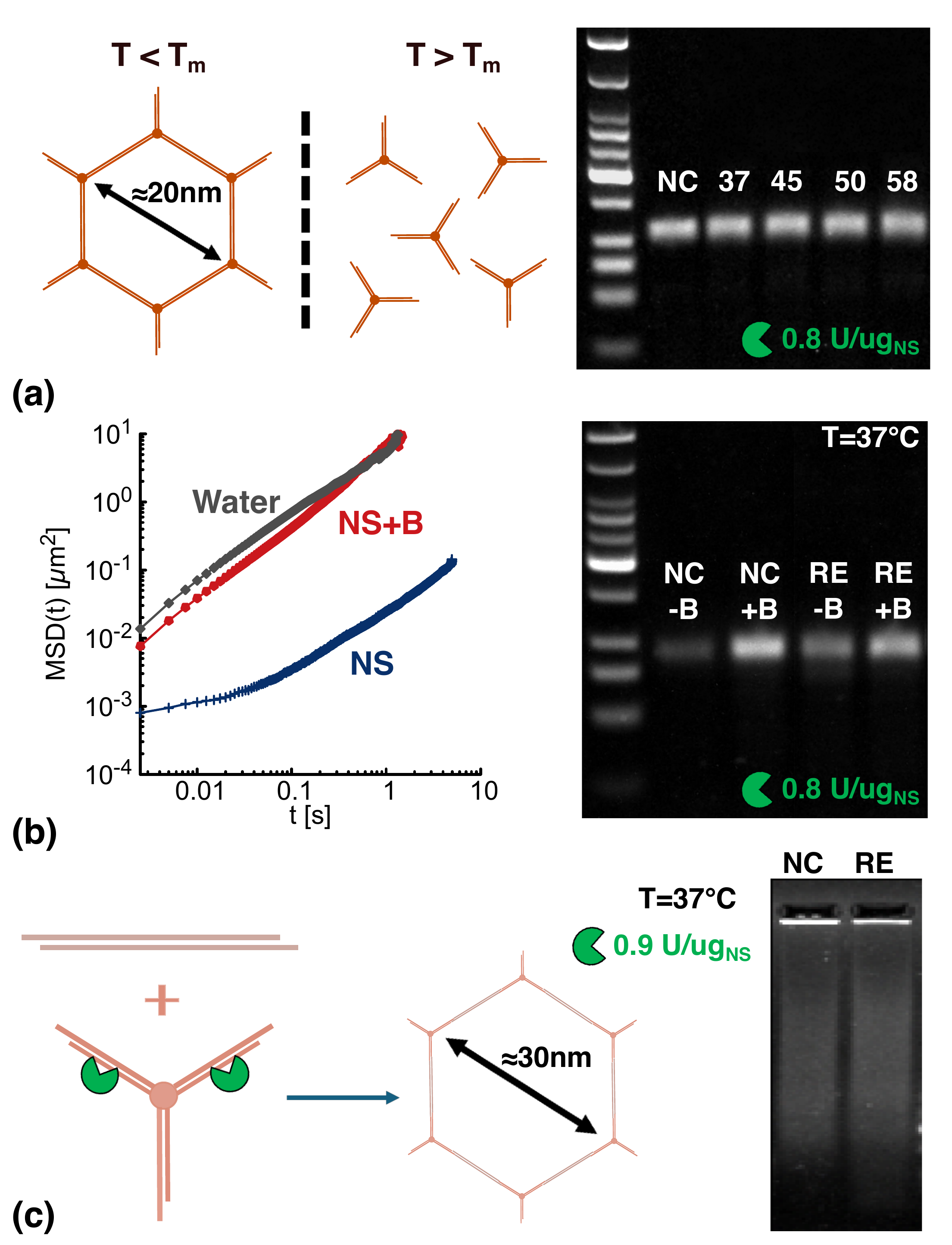}
\caption{\textbf{Hydrogel pore size does not affect EcoRI-mediated degradation}. \textbf{(a)} Left: Schematic of design A DNAns gels below and above the sticky-end melting temperature ($T_{m3}=41^\circ$C). Right: Agarose gel electrophoresis (native conditions) of design A hydrogels incubated with EcoRI for 16h at 37–58$^{\circ}$C. \textbf{(b)} Left: MSD from microrheology for water (gray), design A hydrogel (blue), and hydrogel with ssDNA blockers (red). Right: Agarose gel electrophoresis (native conditions) of design A hydrogels with (+B) and without (-B) blocking strands. All samples were incubated for 16h with EcoRI at 37$^{\circ}$C. Variations in band intensity are attributed to small variations in the loaded sample volume. \textbf{(c)} Left: Schematic of a design B nanostar hydrogel displaying larger pores. Right: Agarose gel electrophoresis (native conditions) of design B hydrogels incubated for 16h with EcoRI at 37$^{\circ}$C and showing no degradation with respect to negative control (NC).
}
\label{fig:Hyp1}
\end{figure}

\begin{figure*}[t!]
\centering
\includegraphics[width=1.0\linewidth]{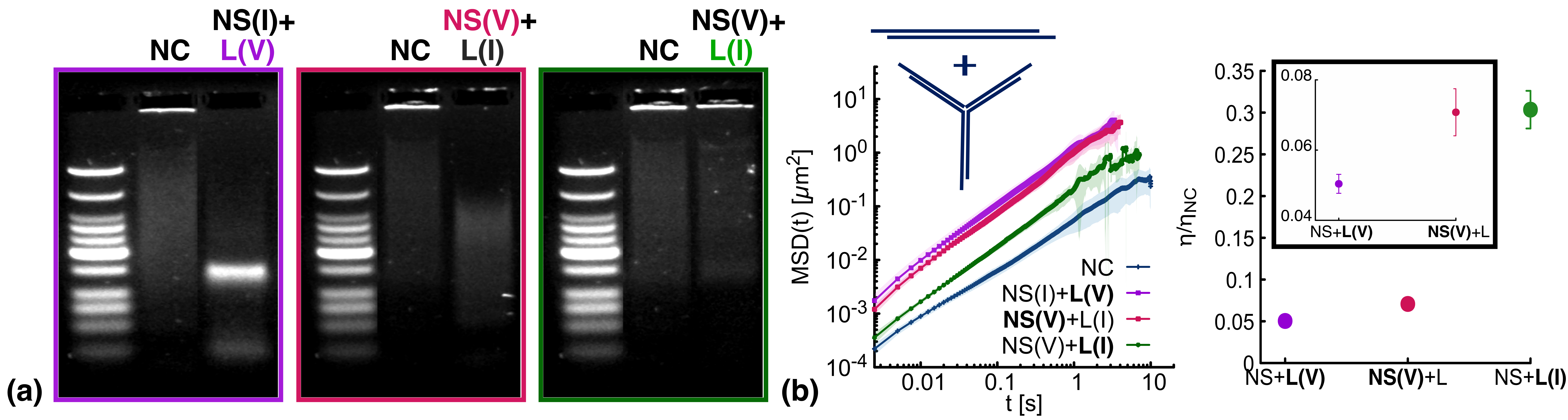}
\caption{\textbf{Effect of site location on DNAns hydrogel cleavage by EcoRI and EcoRV.} \textbf{(a)} Agarose gel electrophoresis (native conditions) after 21h digestion at 37$^\circ$C and 0.68U/$\mu$g. Left: EcoRV digestion of NS(I)+L(V) (site on linker, purple). Middle: EcoRV digestion of NS(V)+L(I) (site on two arms, magenta). Right: EcoRI digestion of NS(V)+L(I) (site on linker, green). NC: samples without enzyme. \textbf{(b)} Left: MSD from microrheology experiments for NC (blue) and digested samples (EcoRV: purple/magenta, EcoRI: green), showing increased mobility after digestion. Right: Normalized viscosity ($\eta/\eta_\mathrm{NC}$) for each case: EcoRV at linker, EcoRV on DNAns arms, EcoRI on linker.
}
\label{fig:Hyp3}
\end{figure*}

To assess whether enzyme degradation efficiency depends on the network’s pore size, we conducted a series of experiments using different DNAns designs while maintaining a constant EcoRI-to-DNA ratio of approximately 0.8U/$\mu$g (100U EcoRI per $\sim$120$\mu$g DNA), corresponding to the highest number of EcoRI units commercially available. We first tested whether varying temperature could enhance EcoRI-mediated degradation of design A nanostar gels. As the temperature approaches the sticky-end melting temperature ($T_{m3}\simeq 41^{\circ}$C), overhang bonds weaken, potentially enlarging the pore size. Above $T_{m3}$, the network is expected to lose connectivity and transition towards a liquid-like state, theoretically improving enzyme diffusion and accessibility. To test this, design A hydrogels were incubated with EcoRI for 16h at fixed temperatures of 37$^\circ$C, 45$^\circ$C, 50$^\circ$C, and 58$^\circ$C. Surprisingly, gel electrophoresis revealed no detectable degradation at any temperature (Fig.~\ref{fig:Hyp1}(a)). This outcome is unexpected, as all tested temperatures lie below the EcoRI inactivation temperature (65$^\circ$C), with 37$^\circ$C corresponding to its optimal activity. These results indicate that increasing pore size alone does not enhance restriction enzyme efficiency in DNA hydrogel degradation.

To further test this conjecture, we introduced oligonucleotides complementary to the sticky ends of design A nanostars in a 3:1 molar ratio. These short strands (6 nucleotides (nt) long) act as blockers, preventing DNAns to bind with each other and in turn disrupting network formation. Microrheology confirmed that DNAns containing blocking strands exhibited water-like viscosity and no elasticity (Fig.~\ref{fig:Hyp1}(b), left). However, gel electrophoresis revealed no detectable EcoRI-mediated degradation after 16h of incubation at 37$^{\circ}$C, regardless of the presence of blocking strands (Fig.~\ref{fig:Hyp1}(b), right).

Finally, we investigated the digestion of a DNAns design that includes a double stranded DNA linker between DNAns, hence enlarging the pore size of the mesh. More specifically, DNAns design B retains the same arm length, flexibility, and EcoRI site placement as design A, but it includes a dsDNA linker that extends the core-to-core contour length distance by $\sim$10 nm.
This distance (comprising one nanostar arm, sticky end, the linker, and another arm) totals 88~bp, corresponding to $\sim$30~nm (assuming 0.34~nm per bp). Also, the longer sticky-ends (8 nt in design B vs 6 nt in design A) form highly stable, larger structures that remain trapped in the well in gel electrophoresis (see negative control (NC) in Fig.~\ref{fig:Hyp1}(c)). Despite these modifications, no detectable digestion was observed after 21 hours of incubation at 37$^\circ$C at 0.9U/$\mu$g (0.45 U/$\mu$g when accounting for both nanostars and linkers), with the DNAns structures remaining in the well, similar to the control.

In conclusion, these experiments collectively indicate that pore size and fluidity of the DNAns network does not determine the efficiency of site-specific REs digestion in DNAns hydrogels. Although the enzymes can diffuse through the network, they still fail to locate and cleave their recognition sites effectively. We therefore reason that this limited accessibility likely arises from geometric and/or steric hindrance imposed by individual nanostar structures, hindering in this way enzyme activity, similarly to the enhanced resistance to enzymatic digestion previously observed in DNA tetrahedral nanostructures~\cite{Bermudez2009}.

\subsection{Relocating restriction site to the linker enhances enzymatic cleavage}
To investigate whether the location of restriction sites affects cleavage efficiency, we compared two restriction enzymes: EcoRI, which upon cleavage produces additional sticky ends, and EcoRV, which generates blunt ends. Both enzymes locate their target sequences by sliding along the DNA, but differ in recognition and cleavage mechanisms~\cite{Pingoud2001}: EcoRI relies on precise hydrogen bonding, whereas EcoRV bends the DNA by $\sim50^\circ$, allowing cleavage \textit{via} conformational changes. These differences are particularly relevant in crowded environments, such as DNAns hydrogels, where steric accessibility can limit enzymatic activity.

\begin{figure*}
\centering
\includegraphics[width=0.85\linewidth]{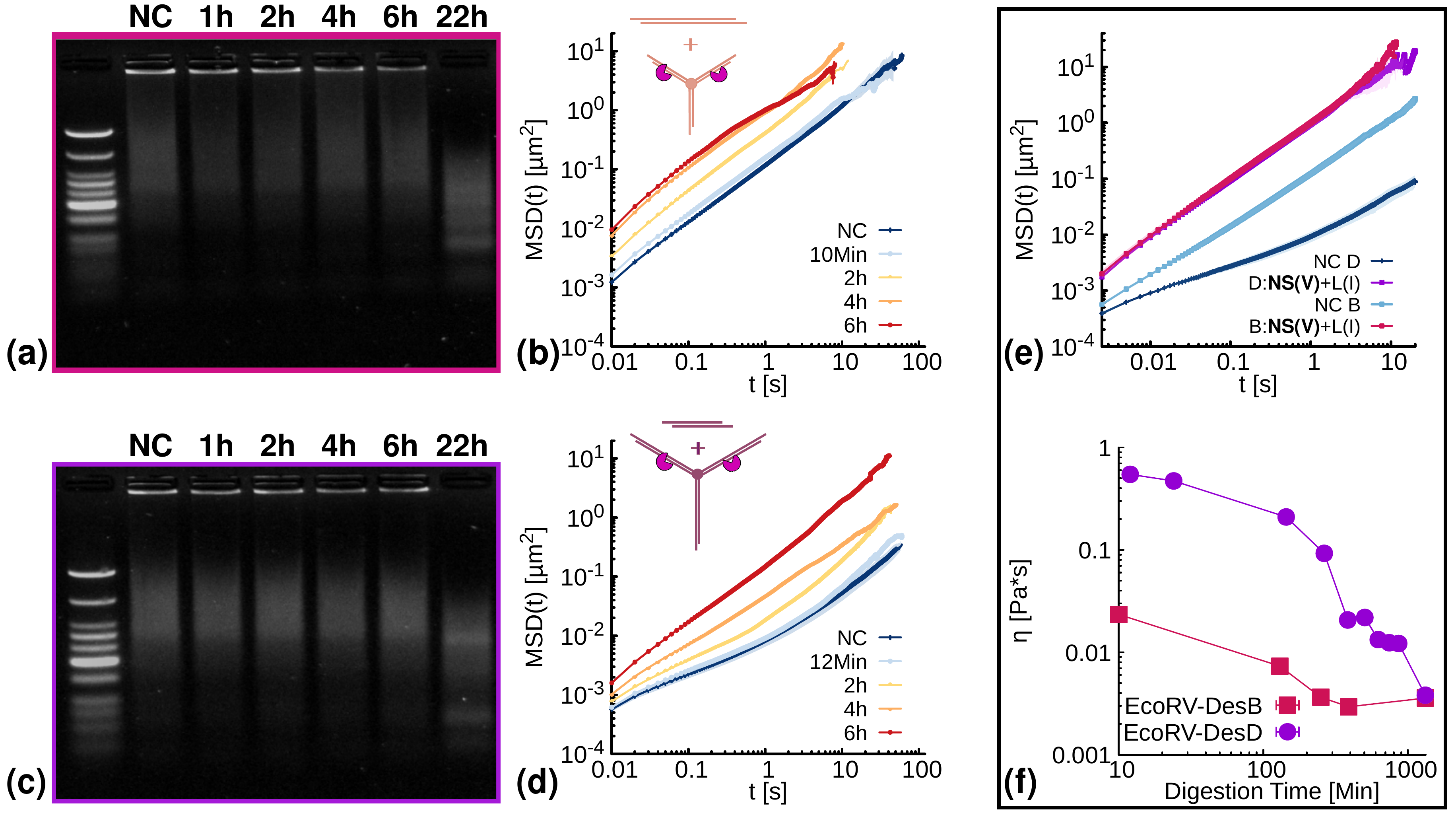}
\caption{\textbf{EcoRV digestion of DNAns designs B and D}. \textbf{(a)} Agarose gel electrophoresis (native conditions) of design B hydrogels digested by EcoRV (100U, at 0.9U/$\mu$g) at 37$^\circ$C. Reactions were stopped at different time points (1h, 2h, 4h, 6h, and 22h). \textbf{(b)} MSD from MR experiments of 500 nm beads over the first 6h of digestion at 37$^\circ$C. \textbf{(c)-(d)} show analogous results as (a)-(b) for design D hydrogels. \textbf{(e)} MSD curves comparing NC with samples digested for 22 h for both designs. \textbf{(f)} Viscosity as function of digestion time.}
\label{fig:ViscosityDesignBD}
\end{figure*}

We work with two versions of design C DNAns: version (i) displays EcoRI sites on two nanostar arms and one EcoRV site on the linker (NS(I)+L(V)), and version (ii) obtained by swapping the EcoRI and EcoRV sites from the first version (NS(V)+L(I)). Compared to design B, design C lacks a flexible core. We hypothesise that if the reason for EcoRI failing to cleave DNAns design B was steric and/or geometric hindrance, then placing the restriction sites on the linker should minimizes these constraints and increase the likelihood of efficient cleavage.

Consistent with this hypothesis, gel electrophoresis and microrheology measurements (at 0.68U/$\mu$g, 100U enzyme in 148$\mu$g mass of linker and DNAns) reported in Fig.~\ref{fig:Hyp3}(a,b), show that EcoRV cleaves most efficiently when the recognition site is on the linker, while partial cleavage occurs for EcoRV at the arms. Furthermore, we observe minimal cleavage for EcoRI at the linker. The effect of EcoRI can best be seen in microrheology (Fig.~\ref{fig:Hyp3}(b)), where the NS(V)+\textbf{L(I)} sample displays lower viscosity than the control but not as low as when the DNAns design C is cut with EcoRV. 

These results highlight the critical role of enzyme choice and restriction site placement for optimizing DNA hydrogel degradation. Relocating the restriction site to the linker (thus farther from the nanostar core) and using the blunt-end enzyme EcoRV (instead of EcoRI) enhances digestion efficiency.

\begin{figure*}[t!]
\centering
\includegraphics[width=0.95\linewidth]{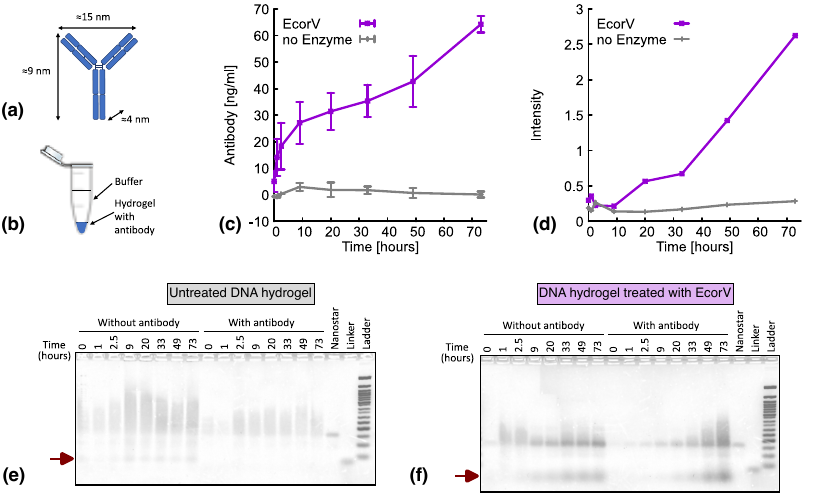}
\caption{\textbf{Antibody encapsulation, retention, and enzyme-mediated release in design D hydrogels}. \textbf{(a)} Schematic of IgG antibody dimensions~\cite{Yih2008}. \textbf{(b)} Antibodies are pre-mixed with DNAns hydrogel. Buffer is added on top of hydrogel sample to measure release over time. \textbf{(c)} Antibody release profiles measured by ELISA for hydrogels treated with EcoRV or no enzyme. \textbf{(d)} Quantification of DNA digestion products in the presence of antibody from the lowest-molecular-weight band in agarose gel electrophoresis (native conditions). \textbf{(e-f)} Representative gel images of hydrogels with and without antibody following treatment with no enzyme and EcoRV, respectively. The arrow indicates the position in the gel of the lowest molecular weight products.
}
\label{fig:antibody}
\end{figure*}

\subsection{DNAns can be designed to encode hydrogel viscoelasticity pre/post degradation}
Having identified conditions under which DNAns hydrogels can undergo degradation by site-specific REs, we next study how intrinsic DNAns design features influence the mechanical response and degradation kinetics of the hydrogels. Specifically, we compared the behaviour of hydrogels made with DNAns designs B and D (where we swapped EcoRI and EcoRV recognition sites for each design in Fig.~\ref{fig:model}(b,d)) when digested by EcoRV (100U) at 37$^\circ$C with an enzyme-to-DNA ratio of 0.9U/$\mu$g. Both designs have a flexible core but differ in arm length and linker length (see Fig.~\ref{fig:model}(e)). These architectural differences yield hydrogels with distinct initial viscoelastic properties and provide a means to assess how network geometry influences both baseline mechanics and enzymatic softening.

Gel electrophoresis at multiple time points (1, 2, 4, 6, and 22 hours) revealed limited digestion within the first 6h for both designs, as DNA remained near the wells, but complete degradation occurred after 22h (Fig.~\ref{fig:ViscosityDesignBD}(a,c)). The degradation can be tracked more precisely with microrheology, which showed that even before full digestion, the MSD of tracer particles increased substantially within the first few hours, reflecting progressive network softening during partial cleavage (Fig.~\ref{fig:ViscosityDesignBD}(b,d)).

Interestingly, the two designs displayed markedly different mechanical responses even before digestion. As shown in Fig.~\ref{fig:ViscosityDesignBD}(e), design D (with longer arms) exhibited a near-plateau in the MSD at early times, reflecting a more elastic-like behavior, whereas design B appeared more liquid-like. Viscosity measurements confirmed this difference: design D initially had a viscosity about an order of magnitude higher than design B, but both converged to similar final values after 22h digestion (Fig.~\ref{fig:ViscosityDesignBD}(f)). The normalized viscosity ratios, $\eta_{22h}/\eta_{0}=0.006$ for D and $\eta_{22h}/\eta_{0}=0.13$ for B, show that D undergoes a 21-fold greater viscosity reduction compared to B. This difference is not only due to variations in enzymatic cleavage efficiency, but rather to the distinct initial architectures of the hydrogels. Together, these results reveal that the temporal evolution of viscoelasticity, and thus the effective degradation kinetics, can be programmed through DNAns geometry.

\subsection{Sequence-specific hydrogel degradation enables controlled antibody release}

To investigate the application of sequence-specific hydrogel degradation for controlled cargo release, we employed design D nanostars, which form hydrogels with pore size $\simeq 34$ nm that is large enough to accommodate Bevacizumab (approximately 15 nm $\times$ 9 nm $\times$ 4 nm in size~\cite{Yih2008}, see Fig.~\ref{fig:antibody}(a)). Bevacizumab, an anti-VEGF monoclonal immunoglobulin G (IgG) currently in clinical trials as a treatment for cancer, was chosen as a model antibody. Since design D DNAns alone do not form a hydrogel in the absence of a linker, this allows us to pre-mix IgG with DNAns in the liquid state and subsequently induce gelation by adding the linker. This strategy facilitates antibody encapsulation without the need for heating, vigorous mixing, or vortexing, which could damage either the hydrogel or the cargo.

To assess antibody retention, DNA hydrogels pre-loaded with antibody were incubated without the addition of any restriction enzyme. DNAns buffer was then applied to the surface of each gel (schematically illustrated in Fig.~\ref{fig:antibody}(b)), and the supernatant was collected at specified time points over 48 hours. Antibody concentrations in the collected buffer were quantified by Enzyme-Linked Immunosorbent Assay (ELISA), revealing negligible passive release, with antibody levels not exceeding 2~ng/ml within 48 hours (see Fig.~\ref{SM-fig:elisaNC}). Detailed descriptions of sample preparation and experimental procedures are provided in the Methods section and SI section~\ref{SM-sec:elisa}.

We next investigated whether sequence-specific hydrogel degradation could trigger antibody release. Antibody-loaded hydrogels were incubated with EcoRV or left untreated as controls, each in triplicates. ELISA measurements of the collected supernatant (Fig.~\ref{fig:antibody}(c)) revealed negligible antibody release from untreated hydrogels over the duration of the experiment, whereas treatment with EcoRV resulted in a pronounced increase in antibody release, with concentrations reaching 64~ng/ml after 73 hours. These results demonstrate that restriction enzyme-mediated cleavage of the hydrogel network provides an effective trigger for the controlled release of encapsulated antibodies.

To directly assess hydrogel degradation, supernatant samples collected at different time points, in the presence and absence of antibody, were analyzed by native agarose gel electrophoresis (Figs.~\ref{fig:antibody}(e-f)). The appearance and increasing intensity of the low-molecular-weight DNA band, corresponding to cleavage products released into the supernatant, confirm progressive EcoRV-mediated degradation of the hydrogel network (Fig.~\ref{fig:antibody}(d)). These observations closely mirror the antibody release profiles measured by ELISA (Fig.~\ref{fig:antibody}(c)), supporting the conclusion that sequence-specific cleavage of DNAns hydrogels enables sustained release of encapsulated cargo.

\section{Conclusions}
We have demonstrated that the degradability and mechanical response of DNAns hydrogels can be programmed through a combination of molecular design and sequence-specific enzymatic cleavage. Under the conditions explored in this study, linker-free hydrogels exhibit strong resistance to digestion by restriction endonucleases, irrespective of pore size (Figs.~\ref{fig:MSD_RE}--\ref{fig:Hyp1}), whereas relocating recognition sites to linker strands enables controlled hydrogel softening and degradation (Fig.~\ref{fig:Hyp3}). Moreover, DNAns geometry independently determines the initial viscoelastic properties and modulates the extent of the mechanical response upon enzymatic cleavage (Fig.~\ref{fig:ViscosityDesignBD}). Finally, we show that DNAns hydrogels can retain antibodies and sequence-specific degradation can be harnessed to trigger the sustained release of encapsulated antibodies (Fig.~\ref{fig:antibody}).

Taken together, these results demonstrate that the mechanical properties and degradation kinetics of DNAns hydrogels can be rationally tuned through molecular design, providing a versatile platform for creating DNA-based materials with tunable structural and functional properties for \textit{in vivo} applications, such as controlled antibody and drug release.

\section{Methods}
\subsection{DNA Nanostar design}
We designed four DNA nanostar (DNAns) variants (A–D) with distinct structural features (see Fig.~\ref{fig:model}). Each nanostar is assembled by annealing three ssDNA oligonucleotides (49nt for A, 50nt for B, 48nt for C, and 74nt for D; sequences in Supporting Information, Tables~\ref{SM-tab:DesignA}–\ref{SM-tab:DesignD}). Equimolar amount of oligos were mixed in nanostar buffer ([NaCl]=150mM, [Tris]=40mM, [Acetate]=40mM, [EDTA]=1mM, pH 8.0) and cooled from 90$^\circ$C to room temperature over 4h to yield three-armed nanostars~\cite{Biffi2013}.

Design A forms self-complementary 6-nt sticky ends (5'-CGATCG-3'), allowing any arm to hybridize with a single arm of another nanostar. Unpaired adenines at the core and before the sticky ends provide flexibility at both the central junction and inter-nanostar connection~\cite{Nguyen2017,Stoev2020}.

Designs B–D use linkers to mediate assembly. Each nanostar arm terminates in an 8-nt non-palindromic sticky end (5'-CGATTGAC-3') that hybridizes to a complementary linker~\cite{Xing2018}. Linkers are formed by annealing two ssDNAs in 150mM NaCl. Mixing DNAns and linkers at a 1:1.5 molar ratio produces hydrogels. Unpaired adenines are retained at the core in designs B and D but omitted in C, resulting in a more rigid structure. In designs B-D there are no unpaired adenines before the the sticky ends, resulting in a more rigid bond between the DNAns and the linker.

Melting temperatures of the nanostars ($T_{m1}$), linkers ($T_{m2}$), and sticky ends ($T_{m3}$) were computed using NUPACK~\cite{nupack} under digestion conditions ([NaCl]=150mM, [Mg$^{2+}$]=10mM, [ssDNA]=250$\mu$M), ensuring $T_{m1} > T_{m3}$ and $T_{m2} > T_{m3}$ by design (Fig.~\ref{fig:model}e).

\subsection{Restriction Enzymes}
To study hydrogel degradation, nanostar arms were engineered to include sequences recognized by REs, illustrated as ``pac-man'' symbols in Fig.~\ref{fig:model} (EcoRI-HF in green, BtgI in red, MspA1I in cyan, and EcoRV-HF in magenta; ``HF'' denotes high-fidelity, omitted for brevity in the main text).

In design A, two arms carry EcoRI sites, while the third contains one BtgI and one MspA1I site (Fig.~\ref{fig:model}a). Designs B–D use EcoRI and EcoRV in complementary configurations: if one enzyme site is on the nanostar, the other is placed on the linker (Fig.~\ref{fig:model}b–d). Full oligo sequences are provided in the Supporting Information.

The enzymes used fall into three categories: (i) sticky-end cutters (EcoRI-HF, BtgI) that generate 4-nt overhangs; (ii) blunt cutters (MspA1I, EcoRV-HF) that leave no overhangs; and (iii) the non-specific endonuclease DNaseI. All enzymes were obtained from New England Biolabs (NEB), which defines 1 unit (U)  as the amount of enzyme required to digest $1\mu$g of substrate DNA at $37^{\circ}$C in a $50\mu$l reaction within 1 h (10 min for DNaseI). For comparison, in our microrheology experiments, we use $\sim100\mu$g of DNA in a $10\mu$l reaction at $37^{\circ}$C.

In all designs, the melting temperature of sticky-end hybridization exceeds the enzymes’ optimal working temperature ((37$^{\circ}$), ensuring network integrity under digestion conditions. Reported enzyme half-lives at (37$^{\circ}$ are 2–4 h (BtgI, EcoRV-HF), 4–8 h (MspA1I), and > 8 h (EcoRI-HF)~\cite{NEBactivity}.

\subsection{Gel electrophoresis}
Gel electrophoresis was performed on 3\% agarose gels pre-stained with SYBR Safe (1:10,000 dilution from stock). For design A, gels were prepared by dissolving 3g agarose (Sigma-Aldrich) in 100mL of 1x LAB buffer (Lithium Acetate Borate). For designs B–D, 1.5g agarose was dissolved in 50mL of 1x TAE buffer (Tris, Acetate, EDTA). Positive and negative controls, as well as reaction samples, were prepared following standard protocols (see Supporting Information, sections~\ref{SM-sec:samplePrep} and \ref{SM-sec:gelelectrophoresis}).

Gel electrophoresis was carried out at room temperature using 150V ($\sim7.5$V/cm using a tray with 20 cm between electrodes) for 35 min for design A and 50 V ($\sim3.3$V/cm) for 70 min for designs B–D. DNA bands were visualized under UV illumination (Syngene GBox) and compared against a low molecular weight DNA ladder (NEB).

\subsection{Particle Tracking Microrheology (PTMR)}
Unless stated otherwise, microrheology experiments were performed at a DNAns concentration of 250$\mu$M (12.2mg/mL DNA) in 10$\mu$L total volume. Sample preparation followed the procedure described in the Gel Electrophoresis: Sample Preparation section of the Supporting Information.

Polystyrene tracer beads (200nm or 500nm diameter, Sigma-Aldrich) were mixed in the sample. To minimize drift and evaporation, $\sim100\mu$m sticky spacers were used to seal the sample between a glass slide and coverslip. Solutions were equilibrated for 5 min at room temperature in a stage-top temperature-controlled chamber (OKO Lab) mounted on an inverted Nikon microscope equipped with a 100x oil-immersion objective (bright-field mode). Two acquisition modes were employed: (i) low-frame-rate tracking of 20–30 beads within a 1024x1024 px field of view at 100 fps for 2 min, repeated every 10 min to monitor temporal evolution; and (ii) high-frame-rate tracking of 5–10 beads within a 512x512 px field at 400 fps for 10 s, also repeated every 10 min.

Particle trajectories were extracted using trackpy (\url{github. com/soft-matter/trackpy}) and analyzed with custom C++ scripts to compute mean-squared displacements. The diffusion coefficient, $D_p$, was obtained from the linear regime of the MSD ($\mathrm{MSD}=2dD_pt$), where $d$ is the dimensionality. The sample viscosity, $\eta$, was then calculated using the Stokes–Einstein relation, $\eta = k_B T/(3\pi D_p a)$ where $k_B$ is the Boltzmann constant, $T$ the absolute temperature, $a$ the tracer radius, and $D_p$ the diffusion coefficient derived from MSD analysis.

\subsection{ELISA}
Antibody release was analysed with an enzyme-linked immunosorbent assay (ELISA) for quantitative detection of human IgG using a Human IgG (Total) Uncoated ELISA Kit (Invitrogen, 88-50550) in accordance with the manufacturer’s instructions. Bevacizumab (BioRad, MCA6089) was used as the human IgG standard. Absorbance at 450 nm was measured with a Tecan Infinite® M200 with i-controlTM software, v2.0, hosted by the Edinburgh Genome Foundry. For details of samples preparation see Supporting Information, section~\ref{SM-sec:elisa}.

\section*{Conflicts of interest}
There are no conflicts to declare.

\section*{Acknowledgements}
YAGF acknowledges support from the Physics of Life, UKRI/\allowbreak Wellcome (grant number EP/T022000/1–\allowbreak PoLNET3). DM acknowledges the Royal Society and the European Research Council (grant agreement No 947918, TAP) for funding. SJR acknowledges BBSRC grant BB/Y008545/1 for funding. The authors also acknowledge the contribution of the COST Action Eutopia, CA17139. We acknowledge technical assistance from Dr. Elliott Chapman from the Edinburgh Genome Foundry, an engineering biology research facility specialising in the modular, automated assembly of DNA constructs and assay development at the University of Edinburgh. For the purpose of open access, we have applied a Creative Commons Attribution (CCBY) license to any author accepted manuscript version arising from this submission.

\bibliography{bibliography}

\begin{thebibliography}{18}%
\makeatletter
\providecommand \@ifxundefined [1]{%
 \@ifx{#1\undefined}
}%
\providecommand \@ifnum [1]{%
 \ifnum #1\expandafter \@firstoftwo
 \else \expandafter \@secondoftwo
 \fi
}%
\providecommand \@ifx [1]{%
 \ifx #1\expandafter \@firstoftwo
 \else \expandafter \@secondoftwo
 \fi
}%
\providecommand \natexlab [1]{#1}%
\providecommand \enquote  [1]{``#1''}%
\providecommand \bibnamefont  [1]{#1}%
\providecommand \bibfnamefont [1]{#1}%
\providecommand \citenamefont [1]{#1}%
\providecommand \href@noop [0]{\@secondoftwo}%
\providecommand \href [0]{\begingroup \@sanitize@url \@href}%
\providecommand \@href[1]{\@@startlink{#1}\@@href}%
\providecommand \@@href[1]{\endgroup#1\@@endlink}%
\providecommand \@sanitize@url [0]{\catcode `\\12\catcode `\$12\catcode
  `\&12\catcode `\#12\catcode `\^12\catcode `\_12\catcode `\%12\relax}%
\providecommand \@@startlink[1]{}%
\providecommand \@@endlink[0]{}%
\providecommand \url  [0]{\begingroup\@sanitize@url \@url }%
\providecommand \@url [1]{\endgroup\@href {#1}{\urlprefix }}%
\providecommand \urlprefix  [0]{URL }%
\providecommand \Eprint [0]{\href }%
\providecommand \doibase [0]{https://doi.org/}%
\providecommand \selectlanguage [0]{\@gobble}%
\providecommand \bibinfo  [0]{\@secondoftwo}%
\providecommand \bibfield  [0]{\@secondoftwo}%
\providecommand \translation [1]{[#1]}%
\providecommand \BibitemOpen [0]{}%
\providecommand \bibitemStop [0]{}%
\providecommand \bibitemNoStop [0]{.\EOS\space}%
\providecommand \EOS [0]{\spacefactor3000\relax}%
\providecommand \BibitemShut  [1]{\csname bibitem#1\endcsname}%
\let\auto@bib@innerbib\@empty
\bibitem [{\citenamefont {Zadeh}\ \emph {et~al.}(2011)\citenamefont {Zadeh},
  \citenamefont {Steenberg}, \citenamefont {Bois}, \citenamefont {Wolfe},
  \citenamefont {Pierce}, \citenamefont {Khan}, \citenamefont {Dirks},\ and\
  \citenamefont {Pierce}}]{nupack}%
  \BibitemOpen
  \bibfield  {author} {\bibinfo {author} {\bibfnamefont {J.~N.}\ \bibnamefont
  {Zadeh}}, \bibinfo {author} {\bibfnamefont {C.~D.}\ \bibnamefont
  {Steenberg}}, \bibinfo {author} {\bibfnamefont {J.~S.}\ \bibnamefont {Bois}},
  \bibinfo {author} {\bibfnamefont {B.~R.}\ \bibnamefont {Wolfe}}, \bibinfo
  {author} {\bibfnamefont {M.~B.}\ \bibnamefont {Pierce}}, \bibinfo {author}
  {\bibfnamefont {A.~R.}\ \bibnamefont {Khan}}, \bibinfo {author}
  {\bibfnamefont {R.~M.}\ \bibnamefont {Dirks}},\ and\ \bibinfo {author}
  {\bibfnamefont {N.~A.}\ \bibnamefont {Pierce}},\ }\href
  {https://doi.org/https://doi.org/10.1002/jcc.21596} {\bibfield  {journal}
  {\bibinfo  {journal} {Journal of Computational Chemistry}\ }\textbf {\bibinfo
  {volume} {32}},\ \bibinfo {pages} {170} (\bibinfo {year} {2011})},\ \Eprint
  {https://arxiv.org/abs/https://onlinelibrary.wiley.com/doi/pdf/10.1002/jcc.21596}
  {https://onlinelibrary.wiley.com/doi/pdf/10.1002/jcc.21596} \BibitemShut
  {NoStop}%
\bibitem [{\citenamefont {Nguyen}\ and\ \citenamefont
  {Saleh}(2017)}]{Nguyen2017}%
  \BibitemOpen
  \bibfield  {author} {\bibinfo {author} {\bibfnamefont {D.~T.}\ \bibnamefont
  {Nguyen}}\ and\ \bibinfo {author} {\bibfnamefont {O.~A.}\ \bibnamefont
  {Saleh}},\ }\href {https://doi.org/10.1039/c7sm00557a} {\bibfield  {journal}
  {\bibinfo  {journal} {Soft Matter}\ }\textbf {\bibinfo {volume} {13}},\
  \bibinfo {pages} {5421} (\bibinfo {year} {2017})}\BibitemShut {NoStop}%
\bibitem [{\citenamefont {Stoev}\ \emph {et~al.}(2020)\citenamefont {Stoev},
  \citenamefont {Cao}, \citenamefont {Caciagli}, \citenamefont {Yu},
  \citenamefont {Ness}, \citenamefont {Liu}, \citenamefont {Ghosh},
  \citenamefont {O’Neill}, \citenamefont {Liu},\ and\ \citenamefont
  {Eiser}}]{Stoev2020}%
  \BibitemOpen
  \bibfield  {author} {\bibinfo {author} {\bibfnamefont {I.~D.}\ \bibnamefont
  {Stoev}}, \bibinfo {author} {\bibfnamefont {T.}~\bibnamefont {Cao}}, \bibinfo
  {author} {\bibfnamefont {A.}~\bibnamefont {Caciagli}}, \bibinfo {author}
  {\bibfnamefont {J.}~\bibnamefont {Yu}}, \bibinfo {author} {\bibfnamefont
  {C.}~\bibnamefont {Ness}}, \bibinfo {author} {\bibfnamefont {R.}~\bibnamefont
  {Liu}}, \bibinfo {author} {\bibfnamefont {R.}~\bibnamefont {Ghosh}}, \bibinfo
  {author} {\bibfnamefont {T.}~\bibnamefont {O’Neill}}, \bibinfo {author}
  {\bibfnamefont {D.}~\bibnamefont {Liu}},\ and\ \bibinfo {author}
  {\bibfnamefont {E.}~\bibnamefont {Eiser}},\ }\href@noop {} {\bibfield
  {journal} {\bibinfo  {journal} {Soft Matter}\ }\textbf {\bibinfo {volume}
  {16}},\ \bibinfo {pages} {990} (\bibinfo {year} {2020})}\BibitemShut
  {NoStop}%
\bibitem [{\citenamefont {Xing}\ \emph {et~al.}(2011)\citenamefont {Xing},
  \citenamefont {Cheng}, \citenamefont {Yang}, \citenamefont {Chen},
  \citenamefont {Zhang}, \citenamefont {Sun}, \citenamefont {Yang},\ and\
  \citenamefont {Liu}}]{Xing2011}%
  \BibitemOpen
  \bibfield  {author} {\bibinfo {author} {\bibfnamefont {Y.}~\bibnamefont
  {Xing}}, \bibinfo {author} {\bibfnamefont {E.}~\bibnamefont {Cheng}},
  \bibinfo {author} {\bibfnamefont {Y.}~\bibnamefont {Yang}}, \bibinfo {author}
  {\bibfnamefont {P.}~\bibnamefont {Chen}}, \bibinfo {author} {\bibfnamefont
  {T.}~\bibnamefont {Zhang}}, \bibinfo {author} {\bibfnamefont
  {Y.}~\bibnamefont {Sun}}, \bibinfo {author} {\bibfnamefont {Z.}~\bibnamefont
  {Yang}},\ and\ \bibinfo {author} {\bibfnamefont {D.}~\bibnamefont {Liu}},\
  }\href@noop {} {\bibfield  {journal} {\bibinfo  {journal} {Advanced
  Materials}\ }\textbf {\bibinfo {volume} {23}},\ \bibinfo {pages} {1117}
  (\bibinfo {year} {2011})}\BibitemShut {NoStop}%
\bibitem [{\citenamefont {Conrad}\ \emph {et~al.}(2019)\citenamefont {Conrad},
  \citenamefont {Kennedy}, \citenamefont {Fygenson},\ and\ \citenamefont
  {Saleh}}]{Conrad2019}%
  \BibitemOpen
  \bibfield  {author} {\bibinfo {author} {\bibfnamefont {N.}~\bibnamefont
  {Conrad}}, \bibinfo {author} {\bibfnamefont {T.}~\bibnamefont {Kennedy}},
  \bibinfo {author} {\bibfnamefont {D.~K.}\ \bibnamefont {Fygenson}},\ and\
  \bibinfo {author} {\bibfnamefont {O.~A.}\ \bibnamefont {Saleh}},\ }\href@noop
  {} {\bibfield  {journal} {\bibinfo  {journal} {Proceedings of the National
  Academy of Sciences}\ }\textbf {\bibinfo {volume} {116}},\ \bibinfo {pages}
  {7238} (\bibinfo {year} {2019})}\BibitemShut {NoStop}%
\bibitem [{\citenamefont {Palombo}\ \emph {et~al.}(2025)\citenamefont
  {Palombo}, \citenamefont {Weir}, \citenamefont {Michieletto},\ and\
  \citenamefont {Gutiérrez~Fosado}}]{Palombo2025}%
  \BibitemOpen
  \bibfield  {author} {\bibinfo {author} {\bibfnamefont {G.}~\bibnamefont
  {Palombo}}, \bibinfo {author} {\bibfnamefont {S.}~\bibnamefont {Weir}},
  \bibinfo {author} {\bibfnamefont {D.}~\bibnamefont {Michieletto}},\ and\
  \bibinfo {author} {\bibfnamefont {Y.~A.}\ \bibnamefont {Gutiérrez~Fosado}},\
  }\href@noop {} {\bibfield  {journal} {\bibinfo  {journal} {Nature materials}\
  }\textbf {\bibinfo {volume} {24}},\ \bibinfo {pages} {454} (\bibinfo {year}
  {2025})}\BibitemShut {NoStop}%
\bibitem [{\citenamefont {Labs}()}]{NEB}%
  \BibitemOpen
  \bibfield  {author} {\bibinfo {author} {\bibfnamefont {N.~E.}\ \bibnamefont
  {Labs}},\ }\href@noop {} {\bibinfo {title} {Optimizing restriction
  endonuclease reactions.}},\ \bibinfo {howpublished}
  {\url{https://www.neb.com/en/tools-and-resources/usage-guidelines/optimizing-restriction-endonuclease-reactions?srsltid=AfmBOoqTy4hZTfsh2YWp2jLhhf2DmqmkTF_zBMP_j7-K9FbgErYZ1QVH}}\BibitemShut
  {NoStop}%
\bibitem [{\citenamefont {Lee}\ \emph {et~al.}(2012)\citenamefont {Lee},
  \citenamefont {Costumbrado}, \citenamefont {Hsu},\ and\ \citenamefont
  {Kim}}]{Agarosegel}%
  \BibitemOpen
  \bibfield  {author} {\bibinfo {author} {\bibfnamefont {P.~Y.}\ \bibnamefont
  {Lee}}, \bibinfo {author} {\bibfnamefont {J.}~\bibnamefont {Costumbrado}},
  \bibinfo {author} {\bibfnamefont {C.~Y.}\ \bibnamefont {Hsu}},\ and\ \bibinfo
  {author} {\bibfnamefont {Y.~H.}\ \bibnamefont {Kim}},\ }\href@noop {}
  {\bibfield  {journal} {\bibinfo  {journal} {J Vis Exp.}\ }\textbf {\bibinfo
  {volume} {62}},\ \bibinfo {pages} {3923} (\bibinfo {year}
  {2012})}\BibitemShut {NoStop}%
\bibitem [{\citenamefont {Calladine}\ and\ \citenamefont
  {Drew}(1997)}]{Calladine1997}%
  \BibitemOpen
  \bibfield  {author} {\bibinfo {author} {\bibfnamefont {C.~R.}\ \bibnamefont
  {Calladine}}\ and\ \bibinfo {author} {\bibfnamefont {H.}~\bibnamefont
  {Drew}},\ }\href@noop {} {\emph {\bibinfo {title} {Understanding DNA: The
  Molecule and How It Works.}}}\ (\bibinfo  {publisher} {Academic Press},\
  \bibinfo {year} {1997})\BibitemShut {NoStop}%
\bibitem [{\citenamefont {Bates}\ and\ \citenamefont
  {Maxwell}(2005)}]{Bates2005}%
  \BibitemOpen
  \bibfield  {author} {\bibinfo {author} {\bibfnamefont {A.~D.}\ \bibnamefont
  {Bates}}\ and\ \bibinfo {author} {\bibfnamefont {A.}~\bibnamefont
  {Maxwell}},\ }\href@noop {} {\emph {\bibinfo {title} {DNA Topology.}}}\
  (\bibinfo  {publisher} {Oxford University Press, USA},\ \bibinfo {year}
  {2005})\BibitemShut {NoStop}%
\bibitem [{\citenamefont {Brody}\ \emph {et~al.}(2004)\citenamefont {Brody},
  \citenamefont {Calhoun}, \citenamefont {Gallmeier}, \citenamefont
  {Creavalle},\ and\ \citenamefont {Kern}}]{Brody2004}%
  \BibitemOpen
  \bibfield  {author} {\bibinfo {author} {\bibfnamefont {J.~R.}\ \bibnamefont
  {Brody}}, \bibinfo {author} {\bibfnamefont {E.~S.}\ \bibnamefont {Calhoun}},
  \bibinfo {author} {\bibfnamefont {E.}~\bibnamefont {Gallmeier}}, \bibinfo
  {author} {\bibfnamefont {T.~D.}\ \bibnamefont {Creavalle}},\ and\ \bibinfo
  {author} {\bibfnamefont {S.~E.}\ \bibnamefont {Kern}},\ }\href@noop {}
  {\bibfield  {journal} {\bibinfo  {journal} {BioTechniques}\ }\textbf
  {\bibinfo {volume} {37}},\ \bibinfo {pages} {598} (\bibinfo {year}
  {2004})}\BibitemShut {NoStop}%
\bibitem [{\citenamefont {Elliott}(2020)}]{Elliott2020}%
  \BibitemOpen
  \bibfield  {author} {\bibinfo {author} {\bibfnamefont {A.~D.}\ \bibnamefont
  {Elliott}},\ }\href {https://doi.org/https://doi.org/10.1002/cpcy.68}
  {\bibfield  {journal} {\bibinfo  {journal} {Current Protocols in Cytometry}\
  }\textbf {\bibinfo {volume} {92}},\ \bibinfo {pages} {e68} (\bibinfo {year}
  {2020})}\BibitemShut {NoStop}%
\bibitem [{\citenamefont {Saleh}\ \emph {et~al.}(2020)\citenamefont {Saleh},
  \citenamefont {jin Jeon},\ and\ \citenamefont {Liedl}}]{Saleh2020PNAS}%
  \BibitemOpen
  \bibfield  {author} {\bibinfo {author} {\bibfnamefont {O.~A.}\ \bibnamefont
  {Saleh}}, \bibinfo {author} {\bibfnamefont {B.}~\bibnamefont {jin Jeon}},\
  and\ \bibinfo {author} {\bibfnamefont {T.}~\bibnamefont {Liedl}},\ }\href
  {https://doi.org/10.1073/pnas.2001654117} {\bibfield  {journal} {\bibinfo
  {journal} {Proceedings of the National Academy of Sciences}\ }\textbf
  {\bibinfo {volume} {117}},\ \bibinfo {pages} {16160} (\bibinfo {year}
  {2020})},\ \Eprint
  {https://arxiv.org/abs/https://www.pnas.org/doi/pdf/10.1073/pnas.2001654117}
  {https://www.pnas.org/doi/pdf/10.1073/pnas.2001654117} \BibitemShut {NoStop}%
\bibitem [{\citenamefont {Jeon}\ \emph {et~al.}(2018)\citenamefont {Jeon},
  \citenamefont {Nguyen}, \citenamefont {Abraham}, \citenamefont {Conrad},
  \citenamefont {Fygenson},\ and\ \citenamefont {Saleh}}]{Saleh2018SM}%
  \BibitemOpen
  \bibfield  {author} {\bibinfo {author} {\bibfnamefont {B.-j.}\ \bibnamefont
  {Jeon}}, \bibinfo {author} {\bibfnamefont {D.~T.}\ \bibnamefont {Nguyen}},
  \bibinfo {author} {\bibfnamefont {G.~R.}\ \bibnamefont {Abraham}}, \bibinfo
  {author} {\bibfnamefont {N.}~\bibnamefont {Conrad}}, \bibinfo {author}
  {\bibfnamefont {D.~K.}\ \bibnamefont {Fygenson}},\ and\ \bibinfo {author}
  {\bibfnamefont {O.~A.}\ \bibnamefont {Saleh}},\ }\href
  {https://doi.org/10.1039/C8SM01085D} {\bibfield  {journal} {\bibinfo
  {journal} {Soft Matter}\ }\textbf {\bibinfo {volume} {14}},\ \bibinfo {pages}
  {7009} (\bibinfo {year} {2018})}\BibitemShut {NoStop}%
\bibitem [{\citenamefont {Liang}\ \emph {et~al.}(2001)\citenamefont {Liang},
  \citenamefont {Song}, \citenamefont {Chen},\ and\ \citenamefont
  {Chu}}]{Glycerol}%
  \BibitemOpen
  \bibfield  {author} {\bibinfo {author} {\bibfnamefont {D.}~\bibnamefont
  {Liang}}, \bibinfo {author} {\bibfnamefont {L.}~\bibnamefont {Song}},
  \bibinfo {author} {\bibfnamefont {Z.}~\bibnamefont {Chen}},\ and\ \bibinfo
  {author} {\bibfnamefont {B.}~\bibnamefont {Chu}},\ }\href@noop {} {\bibfield
  {journal} {\bibinfo  {journal} {J Chromatogr A.}\ }\textbf {\bibinfo {volume}
  {931}},\ \bibinfo {pages} {163} (\bibinfo {year} {2001})}\BibitemShut
  {NoStop}%
\bibitem [{\citenamefont {Sato}\ \emph {et~al.}(2020)\citenamefont {Sato},
  \citenamefont {Sakamoto},\ and\ \citenamefont {Takinoue}}]{Sato2020}%
  \BibitemOpen
  \bibfield  {author} {\bibinfo {author} {\bibfnamefont {Y.}~\bibnamefont
  {Sato}}, \bibinfo {author} {\bibfnamefont {T.}~\bibnamefont {Sakamoto}},\
  and\ \bibinfo {author} {\bibfnamefont {M.}~\bibnamefont {Takinoue}},\
  }\href@noop {} {\bibfield  {journal} {\bibinfo  {journal} {Science Advances}\
  }\textbf {\bibinfo {volume} {6}},\ \bibinfo {pages} {eaba3471} (\bibinfo
  {year} {2020})}\BibitemShut {NoStop}%
\bibitem [{\citenamefont {Agarwal}\ \emph {et~al.}(2022)\citenamefont
  {Agarwal}, \citenamefont {Osmanovic}, \citenamefont {Klocke},\ and\
  \citenamefont {Franco}}]{Franco2022}%
  \BibitemOpen
  \bibfield  {author} {\bibinfo {author} {\bibfnamefont {S.}~\bibnamefont
  {Agarwal}}, \bibinfo {author} {\bibfnamefont {D.}~\bibnamefont {Osmanovic}},
  \bibinfo {author} {\bibfnamefont {M.~A.}\ \bibnamefont {Klocke}},\ and\
  \bibinfo {author} {\bibfnamefont {E.}~\bibnamefont {Franco}},\ }\href@noop {}
  {\bibfield  {journal} {\bibinfo  {journal} {ACS nano}\ }\textbf {\bibinfo
  {volume} {16}},\ \bibinfo {pages} {11842–11851} (\bibinfo {year}
  {2022})}\BibitemShut {NoStop}%
\bibitem [{\citenamefont {Tan}\ \emph {et~al.}(2008)\citenamefont {Tan},
  \citenamefont {Liu}, \citenamefont {Nolting}, \citenamefont {Go},
  \citenamefont {Gervay-Hague},\ and\ \citenamefont {Liu}}]{Yih2008}%
  \BibitemOpen
  \bibfield  {author} {\bibinfo {author} {\bibfnamefont {Y.~H.}\ \bibnamefont
  {Tan}}, \bibinfo {author} {\bibfnamefont {M.}~\bibnamefont {Liu}}, \bibinfo
  {author} {\bibfnamefont {B.}~\bibnamefont {Nolting}}, \bibinfo {author}
  {\bibfnamefont {J.~G.}\ \bibnamefont {Go}}, \bibinfo {author} {\bibfnamefont
  {J.}~\bibnamefont {Gervay-Hague}},\ and\ \bibinfo {author} {\bibfnamefont
  {G.-y.}\ \bibnamefont {Liu}},\ }\href {https://doi.org/10.1021/nn800508f}
  {\bibfield  {journal} {\bibinfo  {journal} {ACS Nano}\ }\textbf {\bibinfo
  {volume} {2}},\ \bibinfo {pages} {2374} (\bibinfo {year} {2008})}\BibitemShut
  {NoStop}%
\end{thebibliography}%


\begin{thebibliography}{37}%
\makeatletter
\providecommand \@ifxundefined [1]{%
 \@ifx{#1\undefined}
}%
\providecommand \@ifnum [1]{%
 \ifnum #1\expandafter \@firstoftwo
 \else \expandafter \@secondoftwo
 \fi
}%
\providecommand \@ifx [1]{%
 \ifx #1\expandafter \@firstoftwo
 \else \expandafter \@secondoftwo
 \fi
}%
\providecommand \natexlab [1]{#1}%
\providecommand \enquote  [1]{``#1''}%
\providecommand \bibnamefont  [1]{#1}%
\providecommand \bibfnamefont [1]{#1}%
\providecommand \citenamefont [1]{#1}%
\providecommand \href@noop [0]{\@secondoftwo}%
\providecommand \href [0]{\begingroup \@sanitize@url \@href}%
\providecommand \@href[1]{\@@startlink{#1}\@@href}%
\providecommand \@@href[1]{\endgroup#1\@@endlink}%
\providecommand \@sanitize@url [0]{\catcode `\\12\catcode `\$12\catcode
  `\&12\catcode `\#12\catcode `\^12\catcode `\_12\catcode `\%12\relax}%
\providecommand \@@startlink[1]{}%
\providecommand \@@endlink[0]{}%
\providecommand \url  [0]{\begingroup\@sanitize@url \@url }%
\providecommand \@url [1]{\endgroup\@href {#1}{\urlprefix }}%
\providecommand \urlprefix  [0]{URL }%
\providecommand \Eprint [0]{\href }%
\providecommand \doibase [0]{https://doi.org/}%
\providecommand \selectlanguage [0]{\@gobble}%
\providecommand \bibinfo  [0]{\@secondoftwo}%
\providecommand \bibfield  [0]{\@secondoftwo}%
\providecommand \translation [1]{[#1]}%
\providecommand \BibitemOpen [0]{}%
\providecommand \bibitemStop [0]{}%
\providecommand \bibitemNoStop [0]{.\EOS\space}%
\providecommand \EOS [0]{\spacefactor3000\relax}%
\providecommand \BibitemShut  [1]{\csname bibitem#1\endcsname}%
\let\auto@bib@innerbib\@empty
\bibitem [{\citenamefont {Seeman}\ and\ \citenamefont
  {Sleiman}(2017)}]{Seeman2017}%
  \BibitemOpen
  \bibfield  {author} {\bibinfo {author} {\bibfnamefont {N.~C.}\ \bibnamefont
  {Seeman}}\ and\ \bibinfo {author} {\bibfnamefont {H.~F.}\ \bibnamefont
  {Sleiman}},\ }\href {https://doi.org/10.1007/978-3-319-68255-6_193}
  {\bibfield  {journal} {\bibinfo  {journal} {Nature Reviews Materials}\
  }\textbf {\bibinfo {volume} {3}},\ \bibinfo {pages} {1} (\bibinfo {year}
  {2017})}\BibitemShut {NoStop}%
\bibitem [{\citenamefont {Rothemund}(2006)}]{Rothemund2006}%
  \BibitemOpen
  \bibfield  {author} {\bibinfo {author} {\bibfnamefont {P.~W.~K.}\
  \bibnamefont {Rothemund}},\ }\href {https://doi.org/10.1038/nature04586}
  {\bibfield  {journal} {\bibinfo  {journal} {Nature}\ }\textbf {\bibinfo
  {volume} {440}},\ \bibinfo {pages} {297} (\bibinfo {year}
  {2006})}\BibitemShut {NoStop}%
\bibitem [{\citenamefont {Palombo}\ \emph {et~al.}(2025)\citenamefont
  {Palombo}, \citenamefont {Weir}, \citenamefont {Michieletto},\ and\
  \citenamefont {Gutiérrez~Fosado}}]{Palombo2025}%
  \BibitemOpen
  \bibfield  {author} {\bibinfo {author} {\bibfnamefont {G.}~\bibnamefont
  {Palombo}}, \bibinfo {author} {\bibfnamefont {S.}~\bibnamefont {Weir}},
  \bibinfo {author} {\bibfnamefont {D.}~\bibnamefont {Michieletto}},\ and\
  \bibinfo {author} {\bibfnamefont {Y.~A.}\ \bibnamefont {Gutiérrez~Fosado}},\
  }\href@noop {} {\bibfield  {journal} {\bibinfo  {journal} {Nature materials}\
  }\textbf {\bibinfo {volume} {24}},\ \bibinfo {pages} {454} (\bibinfo {year}
  {2025})}\BibitemShut {NoStop}%
\bibitem [{\citenamefont {Biffi}\ \emph {et~al.}(2013)\citenamefont {Biffi},
  \citenamefont {Cerbino}, \citenamefont {Bomboi}, \citenamefont {Paraboschi},
  \citenamefont {Asselta}, \citenamefont {Sciortino},\ and\ \citenamefont
  {Bellini}}]{Biffi2013}%
  \BibitemOpen
  \bibfield  {author} {\bibinfo {author} {\bibfnamefont {S.}~\bibnamefont
  {Biffi}}, \bibinfo {author} {\bibfnamefont {R.}~\bibnamefont {Cerbino}},
  \bibinfo {author} {\bibfnamefont {F.}~\bibnamefont {Bomboi}}, \bibinfo
  {author} {\bibfnamefont {E.~M.}\ \bibnamefont {Paraboschi}}, \bibinfo
  {author} {\bibfnamefont {R.}~\bibnamefont {Asselta}}, \bibinfo {author}
  {\bibfnamefont {F.}~\bibnamefont {Sciortino}},\ and\ \bibinfo {author}
  {\bibfnamefont {T.}~\bibnamefont {Bellini}},\ }\href@noop {} {\bibfield
  {journal} {\bibinfo  {journal} {{P}{N}{A}{S}}\ }\textbf {\bibinfo {volume}
  {110}},\ \bibinfo {pages} {15633–} (\bibinfo {year} {2013})}\BibitemShut
  {NoStop}%
\bibitem [{\citenamefont {Biffi}\ \emph {et~al.}(2015)\citenamefont {Biffi},
  \citenamefont {Cerbino}, \citenamefont {Nava}, \citenamefont {Bomboi},
  \citenamefont {Sciortino},\ and\ \citenamefont {Bellini}}]{Biffi2015}%
  \BibitemOpen
  \bibfield  {author} {\bibinfo {author} {\bibfnamefont {S.}~\bibnamefont
  {Biffi}}, \bibinfo {author} {\bibfnamefont {R.}~\bibnamefont {Cerbino}},
  \bibinfo {author} {\bibfnamefont {G.}~\bibnamefont {Nava}}, \bibinfo {author}
  {\bibfnamefont {F.}~\bibnamefont {Bomboi}}, \bibinfo {author} {\bibfnamefont
  {F.}~\bibnamefont {Sciortino}},\ and\ \bibinfo {author} {\bibfnamefont
  {T.}~\bibnamefont {Bellini}},\ }\href {https://doi.org/10.1039/c4sm02144d}
  {\bibfield  {journal} {\bibinfo  {journal} {Soft Matter}\ }\textbf {\bibinfo
  {volume} {11}},\ \bibinfo {pages} {3132} (\bibinfo {year}
  {2015})}\BibitemShut {NoStop}%
\bibitem [{\citenamefont {Conrad}\ \emph {et~al.}(2019)\citenamefont {Conrad},
  \citenamefont {Kennedy}, \citenamefont {Fygenson},\ and\ \citenamefont
  {Saleh}}]{Conrad2019}%
  \BibitemOpen
  \bibfield  {author} {\bibinfo {author} {\bibfnamefont {N.}~\bibnamefont
  {Conrad}}, \bibinfo {author} {\bibfnamefont {T.}~\bibnamefont {Kennedy}},
  \bibinfo {author} {\bibfnamefont {D.~K.}\ \bibnamefont {Fygenson}},\ and\
  \bibinfo {author} {\bibfnamefont {O.~A.}\ \bibnamefont {Saleh}},\ }\href@noop
  {} {\bibfield  {journal} {\bibinfo  {journal} {Proceedings of the National
  Academy of Sciences}\ }\textbf {\bibinfo {volume} {116}},\ \bibinfo {pages}
  {7238} (\bibinfo {year} {2019})}\BibitemShut {NoStop}%
\bibitem [{\citenamefont {Brady}\ \emph {et~al.}(2018)\citenamefont {Brady},
  \citenamefont {Brooks}, \citenamefont {Foder{`a}}, \citenamefont {Cicuta},\
  and\ \citenamefont {Di~Michele}}]{Brady2018AmphiphilicDNA}%
  \BibitemOpen
  \bibfield  {author} {\bibinfo {author} {\bibfnamefont {R.~A.}\ \bibnamefont
  {Brady}}, \bibinfo {author} {\bibfnamefont {N.~J.}\ \bibnamefont {Brooks}},
  \bibinfo {author} {\bibfnamefont {V.}~\bibnamefont {Foder{`a}}}, \bibinfo
  {author} {\bibfnamefont {P.}~\bibnamefont {Cicuta}},\ and\ \bibinfo {author}
  {\bibfnamefont {L.}~\bibnamefont {Di~Michele}},\ }\href
  {https://doi.org/10.1021/jacs.8b09143} {\bibfield  {journal} {\bibinfo
  {journal} {Journal of the American Chemical Society}\ }\textbf {\bibinfo
  {volume} {140}},\ \bibinfo {pages} {15384} (\bibinfo {year}
  {2018})}\BibitemShut {NoStop}%
\bibitem [{\citenamefont {Gadzekpo}\ \emph {et~al.}(2025)\citenamefont
  {Gadzekpo}, \citenamefont {Oprzeska-Zingrebe}, \citenamefont {Kozlowska},
  \citenamefont {Hilbert},\ and\ \citenamefont {Stoev}}]{Stoev2025}%
  \BibitemOpen
  \bibfield  {author} {\bibinfo {author} {\bibfnamefont {A.}~\bibnamefont
  {Gadzekpo}}, \bibinfo {author} {\bibfnamefont {E.~A.}\ \bibnamefont
  {Oprzeska-Zingrebe}}, \bibinfo {author} {\bibfnamefont {M.}~\bibnamefont
  {Kozlowska}}, \bibinfo {author} {\bibfnamefont {L.}~\bibnamefont {Hilbert}},\
  and\ \bibinfo {author} {\bibfnamefont {I.~D.}\ \bibnamefont {Stoev}},\ }\href
  {https://doi.org/https://doi.org/10.1002/adfm.202519573} {\bibfield
  {journal} {\bibinfo  {journal} {Advanced Functional Materials}\ ,\ \bibinfo
  {pages} {e19573}} (\bibinfo {year} {2025})}\BibitemShut {NoStop}%
\bibitem [{\citenamefont {Yuan}\ \emph {et~al.}(2021)\citenamefont {Yuan},
  \citenamefont {Shao}, \citenamefont {Zhou}, \citenamefont {Liu},
  \citenamefont {Zhu}, \citenamefont {Zhou}, \citenamefont {Dong},
  \citenamefont {Stephanopoulos}, \citenamefont {Gui}, \citenamefont {Yan},\
  and\ \citenamefont {Liu}}]{Yuan2021}%
  \BibitemOpen
  \bibfield  {author} {\bibinfo {author} {\bibfnamefont {T.}~\bibnamefont
  {Yuan}}, \bibinfo {author} {\bibfnamefont {Y.}~\bibnamefont {Shao}}, \bibinfo
  {author} {\bibfnamefont {X.}~\bibnamefont {Zhou}}, \bibinfo {author}
  {\bibfnamefont {Q.}~\bibnamefont {Liu}}, \bibinfo {author} {\bibfnamefont
  {Z.}~\bibnamefont {Zhu}}, \bibinfo {author} {\bibfnamefont {B.}~\bibnamefont
  {Zhou}}, \bibinfo {author} {\bibfnamefont {Y.}~\bibnamefont {Dong}}, \bibinfo
  {author} {\bibfnamefont {N.}~\bibnamefont {Stephanopoulos}}, \bibinfo
  {author} {\bibfnamefont {S.}~\bibnamefont {Gui}}, \bibinfo {author}
  {\bibfnamefont {H.}~\bibnamefont {Yan}},\ and\ \bibinfo {author}
  {\bibfnamefont {D.}~\bibnamefont {Liu}},\ }\href
  {https://doi.org/10.1002/adma.202102428} {\bibfield  {journal} {\bibinfo
  {journal} {Advanced Materials}\ }\textbf {\bibinfo {volume} {2102428}},\
  \bibinfo {pages} {1} (\bibinfo {year} {2021})}\BibitemShut {NoStop}%
\bibitem [{\citenamefont {English}\ \emph {et~al.}(2019)\citenamefont
  {English}, \citenamefont {Soenksen}, \citenamefont {Gayet}, \citenamefont
  {de~Puig}, \citenamefont {Angenent-Mari}, \citenamefont {Mao}, \citenamefont
  {Nguyen},\ and\ \citenamefont {Collins}}]{English2019}%
  \BibitemOpen
  \bibfield  {author} {\bibinfo {author} {\bibfnamefont {M.~A.}\ \bibnamefont
  {English}}, \bibinfo {author} {\bibfnamefont {L.~R.}\ \bibnamefont
  {Soenksen}}, \bibinfo {author} {\bibfnamefont {R.~V.}\ \bibnamefont {Gayet}},
  \bibinfo {author} {\bibfnamefont {H.}~\bibnamefont {de~Puig}}, \bibinfo
  {author} {\bibfnamefont {N.~M.}\ \bibnamefont {Angenent-Mari}}, \bibinfo
  {author} {\bibfnamefont {A.~S.}\ \bibnamefont {Mao}}, \bibinfo {author}
  {\bibfnamefont {P.~Q.}\ \bibnamefont {Nguyen}},\ and\ \bibinfo {author}
  {\bibfnamefont {J.~J.}\ \bibnamefont {Collins}},\ }\href
  {https://doi.org/10.1126/science.aaw5122} {\bibfield  {journal} {\bibinfo
  {journal} {Science (New York, N.Y.)}\ }\textbf {\bibinfo {volume} {365}},\
  \bibinfo {pages} {780} (\bibinfo {year} {2019})}\BibitemShut {NoStop}%
\bibitem [{\citenamefont {Malouf}\ \emph {et~al.}(2023)\citenamefont {Malouf},
  \citenamefont {Tanase}, \citenamefont {Fabrini}, \citenamefont {Brady},
  \citenamefont {Paez-Perez}, \citenamefont {Leathers}, \citenamefont {Booth},\
  and\ \citenamefont {Di~Michele}}]{Malouf2023SculptingDNA}%
  \BibitemOpen
  \bibfield  {author} {\bibinfo {author} {\bibfnamefont {L.}~\bibnamefont
  {Malouf}}, \bibinfo {author} {\bibfnamefont {D.~A.}\ \bibnamefont {Tanase}},
  \bibinfo {author} {\bibfnamefont {G.}~\bibnamefont {Fabrini}}, \bibinfo
  {author} {\bibfnamefont {R.~A.}\ \bibnamefont {Brady}}, \bibinfo {author}
  {\bibfnamefont {M.}~\bibnamefont {Paez-Perez}}, \bibinfo {author}
  {\bibfnamefont {A.}~\bibnamefont {Leathers}}, \bibinfo {author}
  {\bibfnamefont {M.~J.}\ \bibnamefont {Booth}},\ and\ \bibinfo {author}
  {\bibfnamefont {L.}~\bibnamefont {Di~Michele}},\ }\href
  {https://doi.org/10.1016/j.chempr.2023.10.004} {\bibfield  {journal}
  {\bibinfo  {journal} {Chem}\ }\textbf {\bibinfo {volume} {9}},\ \bibinfo
  {pages} {3347} (\bibinfo {year} {2023})}\BibitemShut {NoStop}%
\bibitem [{\citenamefont {Bucci}\ \emph {et~al.}(2024)\citenamefont {Bucci},
  \citenamefont {Malouf}, \citenamefont {Tanase}, \citenamefont {Farag},
  \citenamefont {Lamb}, \citenamefont {Rubio-Sánchez}, \citenamefont
  {Gentile}, \citenamefont {Del~Grosso}, \citenamefont {Kaminski},
  \citenamefont {Di~Michele},\ and\ \citenamefont {Ricci}}]{DiMichele2024JACS}%
  \BibitemOpen
  \bibfield  {author} {\bibinfo {author} {\bibfnamefont {J.}~\bibnamefont
  {Bucci}}, \bibinfo {author} {\bibfnamefont {L.}~\bibnamefont {Malouf}},
  \bibinfo {author} {\bibfnamefont {D.~A.}\ \bibnamefont {Tanase}}, \bibinfo
  {author} {\bibfnamefont {N.}~\bibnamefont {Farag}}, \bibinfo {author}
  {\bibfnamefont {J.~R.}\ \bibnamefont {Lamb}}, \bibinfo {author}
  {\bibfnamefont {R.}~\bibnamefont {Rubio-Sánchez}}, \bibinfo {author}
  {\bibfnamefont {S.}~\bibnamefont {Gentile}}, \bibinfo {author} {\bibfnamefont
  {E.}~\bibnamefont {Del~Grosso}}, \bibinfo {author} {\bibfnamefont {C.~F.}\
  \bibnamefont {Kaminski}}, \bibinfo {author} {\bibfnamefont {L.}~\bibnamefont
  {Di~Michele}},\ and\ \bibinfo {author} {\bibfnamefont {F.}~\bibnamefont
  {Ricci}},\ }\href {https://doi.org/10.1021/jacs.4c08919} {\bibfield
  {journal} {\bibinfo  {journal} {Journal of the American Chemical Society}\
  }\textbf {\bibinfo {volume} {146}},\ \bibinfo {pages} {31529} (\bibinfo
  {year} {2024})}\BibitemShut {NoStop}%
\bibitem [{\citenamefont {Tanase}\ \emph {et~al.}(2025)\citenamefont {Tanase},
  \citenamefont {Osmanović}, \citenamefont {Rubio-Sánchez}, \citenamefont
  {Malouf}, \citenamefont {Franco},\ and\ \citenamefont
  {Di~Michele}}]{DiMichele2025}%
  \BibitemOpen
  \bibfield  {author} {\bibinfo {author} {\bibfnamefont {D.~A.}\ \bibnamefont
  {Tanase}}, \bibinfo {author} {\bibfnamefont {D.}~\bibnamefont {Osmanović}},
  \bibinfo {author} {\bibfnamefont {R.}~\bibnamefont {Rubio-Sánchez}},
  \bibinfo {author} {\bibfnamefont {L.}~\bibnamefont {Malouf}}, \bibinfo
  {author} {\bibfnamefont {E.}~\bibnamefont {Franco}},\ and\ \bibinfo {author}
  {\bibfnamefont {L.}~\bibnamefont {Di~Michele}},\ }\href
  {https://doi.org/https://doi.org/10.1002/advs.202506275} {\bibfield
  {journal} {\bibinfo  {journal} {Advanced Science}\ }\textbf {\bibinfo
  {volume} {12}},\ \bibinfo {pages} {e06275} (\bibinfo {year}
  {2025})}\BibitemShut {NoStop}%
\bibitem [{\citenamefont {Bush}\ \emph {et~al.}(2021)\citenamefont {Bush},
  \citenamefont {Hu},\ and\ \citenamefont {Veneziano}}]{Veneziano2021}%
  \BibitemOpen
  \bibfield  {author} {\bibinfo {author} {\bibfnamefont {J.}~\bibnamefont
  {Bush}}, \bibinfo {author} {\bibfnamefont {C.}~\bibnamefont {Hu}},\ and\
  \bibinfo {author} {\bibfnamefont {R.}~\bibnamefont {Veneziano}},\ }\href@noop
  {} {\bibfield  {journal} {\bibinfo  {journal} {{A}pplied {S}ciences}\
  }\textbf {\bibinfo {volume} {11}},\ \bibinfo {pages} {1} (\bibinfo {year}
  {2021})}\BibitemShut {NoStop}%
\bibitem [{\citenamefont {Fabrini}\ \emph {et~al.}(2022)\citenamefont
  {Fabrini}, \citenamefont {Minard}, \citenamefont {Brady}, \citenamefont
  {Di~Antonio},\ and\ \citenamefont {Di~Michele}}]{Fabrini2022QuadStars}%
  \BibitemOpen
  \bibfield  {author} {\bibinfo {author} {\bibfnamefont {G.}~\bibnamefont
  {Fabrini}}, \bibinfo {author} {\bibfnamefont {A.}~\bibnamefont {Minard}},
  \bibinfo {author} {\bibfnamefont {R.~A.}\ \bibnamefont {Brady}}, \bibinfo
  {author} {\bibfnamefont {M.}~\bibnamefont {Di~Antonio}},\ and\ \bibinfo
  {author} {\bibfnamefont {L.}~\bibnamefont {Di~Michele}},\ }\href
  {https://doi.org/10.1021/acs.nanolett.1c03314} {\bibfield  {journal}
  {\bibinfo  {journal} {Nano Letters}\ }\textbf {\bibinfo {volume} {22}},\
  \bibinfo {pages} {689} (\bibinfo {year} {2022})}\BibitemShut {NoStop}%
\bibitem [{\citenamefont {Um}\ \emph {et~al.}(2006)\citenamefont {Um},
  \citenamefont {Lee}, \citenamefont {Park}, \citenamefont {Kwon},
  \citenamefont {Umbach},\ and\ \citenamefont {Luo}}]{Um2006}%
  \BibitemOpen
  \bibfield  {author} {\bibinfo {author} {\bibfnamefont {S.~H.}\ \bibnamefont
  {Um}}, \bibinfo {author} {\bibfnamefont {J.~B.}\ \bibnamefont {Lee}},
  \bibinfo {author} {\bibfnamefont {N.}~\bibnamefont {Park}}, \bibinfo {author}
  {\bibfnamefont {S.~Y.}\ \bibnamefont {Kwon}}, \bibinfo {author}
  {\bibfnamefont {C.~C.}\ \bibnamefont {Umbach}},\ and\ \bibinfo {author}
  {\bibfnamefont {D.}~\bibnamefont {Luo}},\ }\href
  {https://doi.org/10.1038/nmat1741} {\bibfield  {journal} {\bibinfo  {journal}
  {Nature materials}\ }\textbf {\bibinfo {volume} {5}},\ \bibinfo {pages} {797}
  (\bibinfo {year} {2006})}\BibitemShut {NoStop}%
\bibitem [{\citenamefont {Xing}\ \emph {et~al.}(2018)\citenamefont {Xing},
  \citenamefont {Caciagli}, \citenamefont {Cao}, \citenamefont {Stoev},
  \citenamefont {Zupkauskas}, \citenamefont {O'Neill}, \citenamefont {Wenzel},
  \citenamefont {Lamboll}, \citenamefont {Liu},\ and\ \citenamefont
  {Eiser}}]{Xing2018}%
  \BibitemOpen
  \bibfield  {author} {\bibinfo {author} {\bibfnamefont {Z.}~\bibnamefont
  {Xing}}, \bibinfo {author} {\bibfnamefont {A.}~\bibnamefont {Caciagli}},
  \bibinfo {author} {\bibfnamefont {T.}~\bibnamefont {Cao}}, \bibinfo {author}
  {\bibfnamefont {I.}~\bibnamefont {Stoev}}, \bibinfo {author} {\bibfnamefont
  {M.}~\bibnamefont {Zupkauskas}}, \bibinfo {author} {\bibfnamefont
  {T.}~\bibnamefont {O'Neill}}, \bibinfo {author} {\bibfnamefont
  {T.}~\bibnamefont {Wenzel}}, \bibinfo {author} {\bibfnamefont
  {R.}~\bibnamefont {Lamboll}}, \bibinfo {author} {\bibfnamefont
  {D.}~\bibnamefont {Liu}},\ and\ \bibinfo {author} {\bibfnamefont
  {E.}~\bibnamefont {Eiser}},\ }\href {https://doi.org/10.1073/pnas.1722206115}
  {\bibfield  {journal} {\bibinfo  {journal} {{P}{N}{A}{S}}\ }\textbf {\bibinfo
  {volume} {115}},\ \bibinfo {pages} {8137} (\bibinfo {year} {2018})},\ \Eprint
  {https://arxiv.org/abs/1712.07956} {arXiv:1712.07956} \BibitemShut {NoStop}%
\bibitem [{\citenamefont {Nguyen}\ and\ \citenamefont
  {Saleh}(2017)}]{Nguyen2017}%
  \BibitemOpen
  \bibfield  {author} {\bibinfo {author} {\bibfnamefont {D.~T.}\ \bibnamefont
  {Nguyen}}\ and\ \bibinfo {author} {\bibfnamefont {O.~A.}\ \bibnamefont
  {Saleh}},\ }\href {https://doi.org/10.1039/c7sm00557a} {\bibfield  {journal}
  {\bibinfo  {journal} {Soft Matter}\ }\textbf {\bibinfo {volume} {13}},\
  \bibinfo {pages} {5421} (\bibinfo {year} {2017})}\BibitemShut {NoStop}%
\bibitem [{\citenamefont {Fernandez-Castanon}\ \emph
  {et~al.}(2018)\citenamefont {Fernandez-Castanon}, \citenamefont {Bianchi},
  \citenamefont {Saglimbeni}, \citenamefont {{Di Leonardo}},\ and\
  \citenamefont {Sciortino}}]{Fernandez-Castanon2018}%
  \BibitemOpen
  \bibfield  {author} {\bibinfo {author} {\bibfnamefont {J.}~\bibnamefont
  {Fernandez-Castanon}}, \bibinfo {author} {\bibfnamefont {S.}~\bibnamefont
  {Bianchi}}, \bibinfo {author} {\bibfnamefont {F.}~\bibnamefont {Saglimbeni}},
  \bibinfo {author} {\bibfnamefont {R.}~\bibnamefont {{Di Leonardo}}},\ and\
  \bibinfo {author} {\bibfnamefont {F.}~\bibnamefont {Sciortino}},\ }\href
  {https://doi.org/10.1039/c8sm00751a} {\bibfield  {journal} {\bibinfo
  {journal} {Soft Matter}\ }\textbf {\bibinfo {volume} {14}},\ \bibinfo {pages}
  {6431} (\bibinfo {year} {2018})}\BibitemShut {NoStop}%
\bibitem [{\citenamefont {Gutiérrez~Fosado}(2023)}]{Fosado2023}%
  \BibitemOpen
  \bibfield  {author} {\bibinfo {author} {\bibfnamefont {Y.~A.}\ \bibnamefont
  {Gutiérrez~Fosado}},\ }\href {https://doi.org/10.1039/D2SM00221C} {\bibfield
   {journal} {\bibinfo  {journal} {Soft Matter}\ }\textbf {\bibinfo {volume}
  {19}},\ \bibinfo {pages} {4820} (\bibinfo {year} {2023})}\BibitemShut
  {NoStop}%
\bibitem [{\citenamefont {Locatelli}\ \emph {et~al.}(2017)\citenamefont
  {Locatelli}, \citenamefont {Handle}, \citenamefont {Likos}, \citenamefont
  {Sciortino},\ and\ \citenamefont {Rovigatti}}]{Locatelli2017}%
  \BibitemOpen
  \bibfield  {author} {\bibinfo {author} {\bibfnamefont {E.}~\bibnamefont
  {Locatelli}}, \bibinfo {author} {\bibfnamefont {P.~H.}\ \bibnamefont
  {Handle}}, \bibinfo {author} {\bibfnamefont {C.~N.}\ \bibnamefont {Likos}},
  \bibinfo {author} {\bibfnamefont {F.}~\bibnamefont {Sciortino}},\ and\
  \bibinfo {author} {\bibfnamefont {L.}~\bibnamefont {Rovigatti}},\ }\href
  {https://doi.org/10.1021/acsnano.6b08287} {\bibfield  {journal} {\bibinfo
  {journal} {ACS Nano}\ }\textbf {\bibinfo {volume} {11}},\ \bibinfo {pages}
  {2094} (\bibinfo {year} {2017})}\BibitemShut {NoStop}%
\bibitem [{\citenamefont {Zadeh}\ \emph {et~al.}(2011)\citenamefont {Zadeh},
  \citenamefont {Steenberg}, \citenamefont {Bois}, \citenamefont {Wolfe},
  \citenamefont {Pierce}, \citenamefont {Khan}, \citenamefont {Dirks},\ and\
  \citenamefont {Pierce}}]{nupack}%
  \BibitemOpen
  \bibfield  {author} {\bibinfo {author} {\bibfnamefont {J.~N.}\ \bibnamefont
  {Zadeh}}, \bibinfo {author} {\bibfnamefont {C.~D.}\ \bibnamefont
  {Steenberg}}, \bibinfo {author} {\bibfnamefont {J.~S.}\ \bibnamefont {Bois}},
  \bibinfo {author} {\bibfnamefont {B.~R.}\ \bibnamefont {Wolfe}}, \bibinfo
  {author} {\bibfnamefont {M.~B.}\ \bibnamefont {Pierce}}, \bibinfo {author}
  {\bibfnamefont {A.~R.}\ \bibnamefont {Khan}}, \bibinfo {author}
  {\bibfnamefont {R.~M.}\ \bibnamefont {Dirks}},\ and\ \bibinfo {author}
  {\bibfnamefont {N.~A.}\ \bibnamefont {Pierce}},\ }\href
  {https://doi.org/https://doi.org/10.1002/jcc.21596} {\bibfield  {journal}
  {\bibinfo  {journal} {Journal of Computational Chemistry}\ }\textbf {\bibinfo
  {volume} {32}},\ \bibinfo {pages} {170} (\bibinfo {year} {2011})},\ \Eprint
  {https://arxiv.org/abs/https://onlinelibrary.wiley.com/doi/pdf/10.1002/jcc.21596}
  {https://onlinelibrary.wiley.com/doi/pdf/10.1002/jcc.21596} \BibitemShut
  {NoStop}%
\bibitem [{\citenamefont {Bomboi}\ \emph {et~al.}(2016)\citenamefont {Bomboi},
  \citenamefont {Romano}, \citenamefont {Leo}, \citenamefont
  {Fernandez-Castanon}, \citenamefont {Cerbino}, \citenamefont {Bellini},
  \citenamefont {Bordi}, \citenamefont {Filetici},\ and\ \citenamefont
  {Sciortino}}]{Bomboi2016}%
  \BibitemOpen
  \bibfield  {author} {\bibinfo {author} {\bibfnamefont {F.}~\bibnamefont
  {Bomboi}}, \bibinfo {author} {\bibfnamefont {F.}~\bibnamefont {Romano}},
  \bibinfo {author} {\bibfnamefont {M.}~\bibnamefont {Leo}}, \bibinfo {author}
  {\bibfnamefont {J.}~\bibnamefont {Fernandez-Castanon}}, \bibinfo {author}
  {\bibfnamefont {R.}~\bibnamefont {Cerbino}}, \bibinfo {author} {\bibfnamefont
  {T.}~\bibnamefont {Bellini}}, \bibinfo {author} {\bibfnamefont
  {F.}~\bibnamefont {Bordi}}, \bibinfo {author} {\bibfnamefont
  {P.}~\bibnamefont {Filetici}},\ and\ \bibinfo {author} {\bibfnamefont
  {F.}~\bibnamefont {Sciortino}},\ }\href@noop {} {\bibfield  {journal}
  {\bibinfo  {journal} {Nat. Commun.}\ }\textbf {\bibinfo {volume} {7}}
  (\bibinfo {year} {2016})}\BibitemShut {NoStop}%
\bibitem [{\citenamefont {Stoev}\ \emph {et~al.}(2020)\citenamefont {Stoev},
  \citenamefont {Cao}, \citenamefont {Caciagli}, \citenamefont {Yu},
  \citenamefont {Ness}, \citenamefont {Liu}, \citenamefont {Ghosh},
  \citenamefont {O’Neill}, \citenamefont {Liu},\ and\ \citenamefont
  {Eiser}}]{Stoev2020}%
  \BibitemOpen
  \bibfield  {author} {\bibinfo {author} {\bibfnamefont {I.~D.}\ \bibnamefont
  {Stoev}}, \bibinfo {author} {\bibfnamefont {T.}~\bibnamefont {Cao}}, \bibinfo
  {author} {\bibfnamefont {A.}~\bibnamefont {Caciagli}}, \bibinfo {author}
  {\bibfnamefont {J.}~\bibnamefont {Yu}}, \bibinfo {author} {\bibfnamefont
  {C.}~\bibnamefont {Ness}}, \bibinfo {author} {\bibfnamefont {R.}~\bibnamefont
  {Liu}}, \bibinfo {author} {\bibfnamefont {R.}~\bibnamefont {Ghosh}}, \bibinfo
  {author} {\bibfnamefont {T.}~\bibnamefont {O’Neill}}, \bibinfo {author}
  {\bibfnamefont {D.}~\bibnamefont {Liu}},\ and\ \bibinfo {author}
  {\bibfnamefont {E.}~\bibnamefont {Eiser}},\ }\href@noop {} {\bibfield
  {journal} {\bibinfo  {journal} {Soft Matter}\ }\textbf {\bibinfo {volume}
  {16}},\ \bibinfo {pages} {990} (\bibinfo {year} {2020})}\BibitemShut
  {NoStop}%
\bibitem [{\citenamefont {Michieletto}\ \emph {et~al.}(2022)\citenamefont
  {Michieletto}, \citenamefont {Neill}, \citenamefont {Weir}, \citenamefont
  {Evans}, \citenamefont {Crist}, \citenamefont {Martinez},\ and\ \citenamefont
  {Robertson-Anderson}}]{Michieletto2022natcomm}%
  \BibitemOpen
  \bibfield  {author} {\bibinfo {author} {\bibfnamefont {D.}~\bibnamefont
  {Michieletto}}, \bibinfo {author} {\bibfnamefont {P.}~\bibnamefont {Neill}},
  \bibinfo {author} {\bibfnamefont {S.}~\bibnamefont {Weir}}, \bibinfo {author}
  {\bibfnamefont {D.}~\bibnamefont {Evans}}, \bibinfo {author} {\bibfnamefont
  {N.}~\bibnamefont {Crist}}, \bibinfo {author} {\bibfnamefont {V.~A.}\
  \bibnamefont {Martinez}},\ and\ \bibinfo {author} {\bibfnamefont {R.~M.}\
  \bibnamefont {Robertson-Anderson}},\ }\href
  {https://doi.org/10.1038/s41467-022-31828-w} {\bibfield  {journal} {\bibinfo
  {journal} {Nature Communications}\ }\textbf {\bibinfo {volume} {13}}
  (\bibinfo {year} {2022})}\BibitemShut {NoStop}%
\bibitem [{\citenamefont {Panoukidou}\ \emph {et~al.}(2024)\citenamefont
  {Panoukidou}, \citenamefont {Weir}, \citenamefont {Sorichetti}, \citenamefont
  {Fosado}, \citenamefont {Lenz},\ and\ \citenamefont
  {Michieletto}}]{Panoukidou2022}%
  \BibitemOpen
  \bibfield  {author} {\bibinfo {author} {\bibfnamefont {M.}~\bibnamefont
  {Panoukidou}}, \bibinfo {author} {\bibfnamefont {S.}~\bibnamefont {Weir}},
  \bibinfo {author} {\bibfnamefont {V.}~\bibnamefont {Sorichetti}}, \bibinfo
  {author} {\bibfnamefont {Y.~G.}\ \bibnamefont {Fosado}}, \bibinfo {author}
  {\bibfnamefont {M.}~\bibnamefont {Lenz}},\ and\ \bibinfo {author}
  {\bibfnamefont {D.}~\bibnamefont {Michieletto}},\ }\href
  {https://doi.org/10.1103/PhysRevResearch.6.023189} {\bibfield  {journal}
  {\bibinfo  {journal} {Phys. Rev. Res.}\ }\textbf {\bibinfo {volume} {6}},\
  \bibinfo {pages} {023189} (\bibinfo {year} {2024})}\BibitemShut {NoStop}%
\bibitem [{\citenamefont {Conforto}\ \emph {et~al.}(2024)\citenamefont
  {Conforto}, \citenamefont {Gutierrez~Fosado},\ and\ \citenamefont
  {Michieletto}}]{Conforto2024}%
  \BibitemOpen
  \bibfield  {author} {\bibinfo {author} {\bibfnamefont {F.}~\bibnamefont
  {Conforto}}, \bibinfo {author} {\bibfnamefont {Y.}~\bibnamefont
  {Gutierrez~Fosado}},\ and\ \bibinfo {author} {\bibfnamefont {D.}~\bibnamefont
  {Michieletto}},\ }\href {https://doi.org/10.1103/PhysRevResearch.6.033160}
  {\bibfield  {journal} {\bibinfo  {journal} {Phys. Rev. Res.}\ }\textbf
  {\bibinfo {volume} {6}},\ \bibinfo {pages} {033160} (\bibinfo {year}
  {2024})}\BibitemShut {NoStop}%
\bibitem [{\citenamefont {Fosado}\ \emph {et~al.}(2023)\citenamefont {Fosado},
  \citenamefont {Howard}, \citenamefont {Weir}, \citenamefont {Noy},
  \citenamefont {Leake},\ and\ \citenamefont {Michieletto}}]{Fosado2022}%
  \BibitemOpen
  \bibfield  {author} {\bibinfo {author} {\bibfnamefont {Y.~A.~G.}\
  \bibnamefont {Fosado}}, \bibinfo {author} {\bibfnamefont {J.}~\bibnamefont
  {Howard}}, \bibinfo {author} {\bibfnamefont {S.}~\bibnamefont {Weir}},
  \bibinfo {author} {\bibfnamefont {A.}~\bibnamefont {Noy}}, \bibinfo {author}
  {\bibfnamefont {M.~C.}\ \bibnamefont {Leake}},\ and\ \bibinfo {author}
  {\bibfnamefont {D.}~\bibnamefont {Michieletto}},\ }\href
  {https://doi.org/10.1103/PhysRevLett.130.058203} {\bibfield  {journal}
  {\bibinfo  {journal} {Physical Review Letters}\ }\textbf {\bibinfo {volume}
  {130}},\ \bibinfo {pages} {058203} (\bibinfo {year} {2023})}\BibitemShut
  {NoStop}%
\bibitem [{\citenamefont {Saleh}\ \emph {et~al.}(2020)\citenamefont {Saleh},
  \citenamefont {jin Jeon},\ and\ \citenamefont {Liedl}}]{Saleh2020PNAS}%
  \BibitemOpen
  \bibfield  {author} {\bibinfo {author} {\bibfnamefont {O.~A.}\ \bibnamefont
  {Saleh}}, \bibinfo {author} {\bibfnamefont {B.}~\bibnamefont {jin Jeon}},\
  and\ \bibinfo {author} {\bibfnamefont {T.}~\bibnamefont {Liedl}},\ }\href
  {https://doi.org/10.1073/pnas.2001654117} {\bibfield  {journal} {\bibinfo
  {journal} {Proceedings of the National Academy of Sciences}\ }\textbf
  {\bibinfo {volume} {117}},\ \bibinfo {pages} {16160} (\bibinfo {year}
  {2020})},\ \Eprint
  {https://arxiv.org/abs/https://www.pnas.org/doi/pdf/10.1073/pnas.2001654117}
  {https://www.pnas.org/doi/pdf/10.1073/pnas.2001654117} \BibitemShut {NoStop}%
\bibitem [{\citenamefont {Saleh}\ \emph {et~al.}(2023)\citenamefont {Saleh},
  \citenamefont {Wilken}, \citenamefont {Squires},\ and\ \citenamefont
  {Liedl}}]{Saleh2023}%
  \BibitemOpen
  \bibfield  {author} {\bibinfo {author} {\bibfnamefont {O.~A.}\ \bibnamefont
  {Saleh}}, \bibinfo {author} {\bibfnamefont {S.}~\bibnamefont {Wilken}},
  \bibinfo {author} {\bibfnamefont {T.~M.}\ \bibnamefont {Squires}},\ and\
  \bibinfo {author} {\bibfnamefont {T.}~\bibnamefont {Liedl}},\ }\href
  {https://doi.org/10.1038/s41467-023-39175-0} {\bibfield  {journal} {\bibinfo
  {journal} {Nature Communications}\ }\textbf {\bibinfo {volume} {14}},\
  \bibinfo {pages} {1} (\bibinfo {year} {2023})}\BibitemShut {NoStop}%
\bibitem [{\citenamefont {Xing}\ \emph {et~al.}(2011)\citenamefont {Xing},
  \citenamefont {Cheng}, \citenamefont {Yang}, \citenamefont {Chen},
  \citenamefont {Zhang}, \citenamefont {Sun}, \citenamefont {Yang},\ and\
  \citenamefont {Liu}}]{Xing2011}%
  \BibitemOpen
  \bibfield  {author} {\bibinfo {author} {\bibfnamefont {Y.}~\bibnamefont
  {Xing}}, \bibinfo {author} {\bibfnamefont {E.}~\bibnamefont {Cheng}},
  \bibinfo {author} {\bibfnamefont {Y.}~\bibnamefont {Yang}}, \bibinfo {author}
  {\bibfnamefont {P.}~\bibnamefont {Chen}}, \bibinfo {author} {\bibfnamefont
  {T.}~\bibnamefont {Zhang}}, \bibinfo {author} {\bibfnamefont
  {Y.}~\bibnamefont {Sun}}, \bibinfo {author} {\bibfnamefont {Z.}~\bibnamefont
  {Yang}},\ and\ \bibinfo {author} {\bibfnamefont {D.}~\bibnamefont {Liu}},\
  }\href@noop {} {\bibfield  {journal} {\bibinfo  {journal} {Advanced
  Materials}\ }\textbf {\bibinfo {volume} {23}},\ \bibinfo {pages} {1117}
  (\bibinfo {year} {2011})}\BibitemShut {NoStop}%
\bibitem [{\citenamefont {Kim}\ \emph {et~al.}(1990)\citenamefont {Kim},
  \citenamefont {Grable}, \citenamefont {Love}, \citenamefont {Greene},\ and\
  \citenamefont {Rosenberg}}]{Kim1990}%
  \BibitemOpen
  \bibfield  {author} {\bibinfo {author} {\bibfnamefont {Y.}~\bibnamefont
  {Kim}}, \bibinfo {author} {\bibfnamefont {J.~C.}\ \bibnamefont {Grable}},
  \bibinfo {author} {\bibfnamefont {R.}~\bibnamefont {Love}}, \bibinfo {author}
  {\bibfnamefont {P.~J.}\ \bibnamefont {Greene}},\ and\ \bibinfo {author}
  {\bibfnamefont {J.~M.}\ \bibnamefont {Rosenberg}},\ }\href
  {https://doi.org/10.1126/science.2399465} {\bibfield  {journal} {\bibinfo
  {journal} {Science}\ }\textbf {\bibinfo {volume} {249}},\ \bibinfo {pages}
  {1307} (\bibinfo {year} {1990})}\BibitemShut {NoStop}%
\bibitem [{\citenamefont {Pingoud}\ and\ \citenamefont
  {Jeltsch}(2001)}]{Pingoud2001}%
  \BibitemOpen
  \bibfield  {author} {\bibinfo {author} {\bibfnamefont {A.}~\bibnamefont
  {Pingoud}}\ and\ \bibinfo {author} {\bibfnamefont {A.}~\bibnamefont
  {Jeltsch}},\ }\href {https://doi.org/10.1093/nar/29.18.3705} {\bibfield
  {journal} {\bibinfo  {journal} {Nucleic Acids Research}\ }\textbf {\bibinfo
  {volume} {29}},\ \bibinfo {pages} {3705} (\bibinfo {year}
  {2001})}\BibitemShut {NoStop}%
\bibitem [{\citenamefont {Suck}(1994)}]{Dietrich1994}%
  \BibitemOpen
  \bibfield  {author} {\bibinfo {author} {\bibfnamefont {D.}~\bibnamefont
  {Suck}},\ }\href {https://doi.org/https://doi.org/10.1002/jmr.300070203}
  {\bibfield  {journal} {\bibinfo  {journal} {Journal of Molecular
  Recognition}\ }\textbf {\bibinfo {volume} {7}},\ \bibinfo {pages} {65}
  (\bibinfo {year} {1994})}\BibitemShut {NoStop}%
\bibitem [{\citenamefont {Keum}\ and\ \citenamefont
  {Bermudez}(2009)}]{Bermudez2009}%
  \BibitemOpen
  \bibfield  {author} {\bibinfo {author} {\bibfnamefont {J.-W.}\ \bibnamefont
  {Keum}}\ and\ \bibinfo {author} {\bibfnamefont {H.}~\bibnamefont
  {Bermudez}},\ }\href {https://doi.org/10.1039/B917661F} {\bibfield  {journal}
  {\bibinfo  {journal} {Chem. Commun.}\ ,\ \bibinfo {pages} {7036}} (\bibinfo
  {year} {2009})}\BibitemShut {NoStop}%
\bibitem [{\citenamefont {Tan}\ \emph {et~al.}(2008)\citenamefont {Tan},
  \citenamefont {Liu}, \citenamefont {Nolting}, \citenamefont {Go},
  \citenamefont {Gervay-Hague},\ and\ \citenamefont {Liu}}]{Yih2008}%
  \BibitemOpen
  \bibfield  {author} {\bibinfo {author} {\bibfnamefont {Y.~H.}\ \bibnamefont
  {Tan}}, \bibinfo {author} {\bibfnamefont {M.}~\bibnamefont {Liu}}, \bibinfo
  {author} {\bibfnamefont {B.}~\bibnamefont {Nolting}}, \bibinfo {author}
  {\bibfnamefont {J.~G.}\ \bibnamefont {Go}}, \bibinfo {author} {\bibfnamefont
  {J.}~\bibnamefont {Gervay-Hague}},\ and\ \bibinfo {author} {\bibfnamefont
  {G.-y.}\ \bibnamefont {Liu}},\ }\href {https://doi.org/10.1021/nn800508f}
  {\bibfield  {journal} {\bibinfo  {journal} {ACS Nano}\ }\textbf {\bibinfo
  {volume} {2}},\ \bibinfo {pages} {2374} (\bibinfo {year} {2008})}\BibitemShut
  {NoStop}%
\bibitem [{\citenamefont {NEB}()}]{NEBactivity}%
  \BibitemOpen
  \bibfield  {author} {\bibinfo {author} {\bibnamefont {NEB}},\ }\href@noop {}
  {\bibinfo {title} {New england biolabs, activity of restriction enzymes}},\
  \bibinfo {howpublished}
  {\url{https://international.neb.com/tools-and-resources/usage-guidelines/restriction-endonucleases-survival-in-a-reaction}},\
  \bibinfo {note} {accessed: 2022-11-13}\BibitemShut {NoStop}%
\end{thebibliography}%
\end{document}


\title{Supporting Information: Designing DNA nanostar hydrogels with programmable degradation and antibody release}

\author{Giorgia Palombo}
\affiliation{School of Physics and Astronomy, University of Edinburgh, Peter Guthrie Tait Road, Edinburgh, EH9 3FD, UK}

\author{Christine A. Merrick}
\affiliation{Centre for Engineering Biology, School of Biological Sciences, University of Edinburgh, Edinburgh, UK}

\author{Jennifer Harnett}
\affiliation{School of Physics and Astronomy, University of Edinburgh, Peter Guthrie Tait Road, Edinburgh, EH9 3FD, UK}

\author{Susan Rosser}
\affiliation{Centre for Engineering Biology, School of Biological Sciences, University of Edinburgh, Edinburgh, UK}

\author{Davide Michieletto}
\thanks{corresponding author, davide.michieletto@ed.ac.uk}
\affiliation{School of Physics and Astronomy, University of Edinburgh, Peter Guthrie Tait Road, Edinburgh, EH9 3FD, UK}
\affiliation{MRC Human Genetics Unit, Institute of Genetics and Cancer, University of Edinburgh, UK}
\affiliation{International Institute for Sustainability with Knotted Chiral Meta Matter (WPI-SKCM\textsuperscript{2}), Hiroshima University, Higashi-Hiroshima, Hiroshima 739-8526, Japan}

\author{Yair Augusto Guti\'{e}rrez Fosado}
\thanks{corresponding author, yair.fosado@ed.ac.uk}
\affiliation{School of Physics and Astronomy, University of Edinburgh, Peter Guthrie Tait Road, Edinburgh, EH9 3FD, UK}

\maketitle

\section{Design specifications of DNA nanostars}
\label{sec:designSequences}
As mentioned in the main text, we designed four distinct types of DNAns building blocks using NUPACK~\cite{nupack}. Each DNAns was assembled from three single-stranded DNA (ssDNA) oligonucleotides (oligos). Upon hybridization, these oligos form a three-armed (Y-shaped) structure (see Fig.~\ref{M-fig:model}).

In design A, each arm consists of a 20 base-pair (bp) segment terminating in a self-complementary 6-nucleotide (nt) fragment (5'-CGATCG-3'). This sticky end is identical on all three arms, allowing any arm of one nanostar to hybridize with any (but only one) arm of another nanostar. Unpaired adenines are introduced as flexible joints (FJ) at two positions: at the core of the Y-shaped structure (FJ1) and before the sticky ends (FJ2). These flexible regions enhance the overall flexibility of both the nanostar and its interconnections \cite{Nguyen2017,Stoev2020}. The specific sequences used are provided in Table~\ref{tab:DesignA}, which also include the recognition sites for the restriction enzymes (REs).

For designs B--D, each DNAns arm (20~bp long for B and C, and 32~bp long for D) terminates in a non-self-complementary 8 nt sticky end sequence, 5'-CGATTGAC-3'. This sequence was specifically chosen to prevent self-assembly between DNAns \cite{Xing2011}. Instead, these sticky ends are complementary to those on the linker (L), which contains two overhangs that mediate hybridization between different nanostars. The linker is a linear double-stranded DNA segment whose length varies among designs: 32~bp for designs B and C, and 20~bp for design D.

Additionally, in designs B-D, there are no unpaired adenines before the sticky ends, resulting in a more rigid connection between the DNAns and the linker. However, the unpaired adenines at the core are retained in designs B and D, but omitted in design C, leading to an overall more rigid structure. The sequences corresponding to these designs are detailed in Tables~\ref{tab:DesignB}, \ref{tab:DesignC}, and \ref{tab:DesignD}.

\begin{table*}[ht]
	\centering  
	\resizebox{1.0\textwidth}{!}{%
		\begin{tabular}{|c|c|}
			\hline
			&Design A: NS(\EcoRI{\bf{EcoRI}}, \MspAI{\bf{MspAI}}, \BtgI{\bf{BtgI}}) \\
			\hline
			NS1 & $5^{\prime}-$GTCAATGC\MspAI{\bf{\boxed{\BtgI{CCGCGG}}}}GCAATT/\textbf{AA}/GGGCATG\EcoRI{\bf{GAATTC}}CGCATCC/\textbf{A}/\underline{CGATCG}$-3^{\prime}$\\
			NS2 & $5^{\prime}-$GGATGCG\EcoRI{\bf{GAATTC}}CATGCCC/\textbf{AA}/CCCTTGG\EcoRI{\bf{GAATTC}}GATCCCG/\textbf{A}/\underline{CGATCG}$-3^{\prime}$\\
			NS3 & $5^{\prime}-$CGGGATC\EcoRI{\bf{GAATTC}}CCAAGGG/\textbf{AA}/AATTGC\MspAI{\bf{\boxed{\BtgI{CCGCGG}}}}GCATTGAC/\textbf{A}/\underline{CGATCG}$-3^{\prime}$\\
			\hline
		\end{tabular}
	}
	\caption{\textbf{Design A sequences}: Each oligo is 49 nt long, with a FJ1 at the core (\textbf{/AA/}) and an additional FJ2 motif preceding the sticky ends (\textbf{/A/}). The sticky end is 6 nt long (\underline{CGATCG}). Restriction enzyme sites are indicated as: EcoRI (green), BtgI (red) and MspA1I (blue box).}
	\label{tab:DesignA}
\end{table*}

\vspace{3cm}
\begin{table*}[htbp]
	\centering 
	\resizebox{1.0\textwidth}{!}{
		\begin{tabular}{|c|c|}
			\hline
			&Design B: NS(\EcoRI{\bf{EcoRI}}) + L(\EcoRV{\bf{EcoRV}}) \\
			\hline
			NS1 & $5^{\prime}-$GTCAATGCCCGCGGGCAATT/\textbf{AA}/GGGCATG\EcoRI{\bf{GAATTC}}CGCATCC/\underline{CGATTGAC}$-3^{\prime}$\\
			NS2 & $5^{\prime}-$GGATGCG\EcoRI{\bf{GAATTC}}CATGCCC/\textbf{AA}/CCCTTGG\EcoRI{\bf{GAATTC}}GATCCCG/\underline{CGATTGAC}$-3^{\prime}$\\
			NS3 & $5^{\prime}-$CGGGATC\EcoRI{\bf{GAATTC}}CCAAGGG/\textbf{AA}/AATTGCCCGCGGGCATTGAC/\underline{CGATTGAC}$-3^{\prime}$\\
			L1 & $5^{\prime}-$CCTATTCGCATGA\EcoRV{\bf{GATATC}}CATTCACCGTAAG/\underline{GTCAATCG}$-3^{\prime}$\\
			L2 & $5^{\prime}-$CTTACGGTGAATG\EcoRV{\bf{GATATC}}TCATGCGAATAGG/\underline{GTCAATCG}$-3^{\prime}$\\
			\hline
			&Design B: NS(\EcoRV{\bf{EcoRV}}) + L(\EcoRI{\bf{EcoRI}})  \\
			\hline
			NS1 & $5^{\prime}-$GTCAATGCCCGCGGGCAATT/\textbf{AA}/GGGCATG\EcoRV{\bf{GATATC}}CGCATCC/\underline{CGATTGAC}$-3^{\prime}$\\
			NS2 & $5^{\prime}-$GGATGCG\EcoRV{\bf{GATATC}}CATGCCC/\textbf{AA}/CCCTTGG\EcoRV{\bf{GATATC}}GATCCCG/\underline{CGATTGAC}$-3^{\prime}$\\
			NS3 & $5^{\prime}-$CGGGATC\EcoRV{\bf{GATATC}}CCAAGGG/\textbf{AA}/AATTGCCCGCGGGCATTGAC/\underline{CGATTGAC}$-3^{\prime}$\\
			L1 & $5^{\prime}-$CCTATTCGCATGA\EcoRI{\bf{GAATTC}}CATTCACCGTAAG/\underline{GTCAATCG}$-3^{\prime}$\\
			L2 & $5^{\prime}-$CTTACGGTGAATG\EcoRI{\bf{GAATTC}}TCATGCGAATAGG/\underline{GTCAATCG}$-3^{\prime}$\\
			\hline
		\end{tabular}
	}
	\caption{\textbf{Design B sequences}: Each NS oligo is 50 nt long, with a FJ1 at the core (\textbf{/AA/}). The NS sticky end is 8 nt long (\underline{CGATTGAC}). Linker oligos are 40 nt long, each with an 8 nt sticky end (\underline{GTCAATCG}), which is complementary to that of the NS. RE sites are labelled as follows: EcoRI (green) and EcoRV (magenta). The oligo sequences with the RE sites inverted between NS and L are also reported.}
	\label{tab:DesignB}
\end{table*}

\vspace{2cm}
\begin{table*}[h!]
	\centering 
	\resizebox{1.0\textwidth}{!}{
		\begin{tabular}{|c|c|}
			\hline
			&Design C: NS(\EcoRI{\bf{EcoRI}})+ L(\EcoRV{\bf{EcoRV}}) \\
			\hline
			NS1 & $5^{\prime}-$GTCAATGCCCGCGGGCAATTGGGCATG\EcoRI{\bf{GAATTC}}CGCATCC/\underline{CGATTGAC}$-3^{\prime}$\\
			NS2 & $5^{\prime}-$GGATGCG\EcoRI{\bf{GAATTC}}CATGCCCCCCTTGG\EcoRI{\bf{GAATTC}}GATCCCG/\underline{CGATTGAC}$-3^{\prime}$\\
			NS3 & $5^{\prime}-$CGGGATC\EcoRI{\bf{GAATTC}}CCAAGGGAATTGCCCGCGGGCATTGAC/\underline{CGATTGAC}$-3^{\prime}$\\
			L1 & $5^{\prime}-$CCTATTCGCATGA\EcoRV{\bf{GATATC}}CATTCACCGTAAG/\underline{GTCAATCG}$-3^{\prime}$\\
			L2 & $5^{\prime}-$CTTACGGTGAATG\EcoRV{\bf{GATATC}}TCATGCGAATAGG/\underline{GTCAATCG}$-3^{\prime}$\\
			\hline
			&Design C: NS(\EcoRV{\bf{EcoRV}}) + L(\EcoRI{\bf{EcoRI}})  \\
			\hline
			NS1 & $5^{\prime}-$GTCAATGCCCGCGGGCAATTGGGCATG\EcoRV{\bf{GATATC}}CGCATCC/\underline{CGATTGAC}$-3^{\prime}$\\
			NS2 & $5^{\prime}-$GGATGCG\EcoRV{\bf{GATATC}}CATGCCCCCCTTGG\EcoRV{\bf{GATATC}}GATCCCG/\underline{CGATTGAC}$-3^{\prime}$\\
			NS3 & $5^{\prime}-$CGGGATC\EcoRV{\bf{GATATC}}CCAAGGGAATTGCCCGCGGGCATTGAC/\underline{CGATTGAC}$-3^{\prime}$\\
			L1 & $5^{\prime}-$CCTATTCGCATGA\EcoRI{\bf{GAATTC}}CATTCACCGTAAG/\underline{GTCAATCG}$-3^{\prime}$\\
			L2 & $5^{\prime}-$CTTACGGTGAATG\EcoRI{\bf{GAATTC}}TCATGCGAATAGG/\underline{GTCAATCG}$-3^{\prime}$\\
			\hline
		\end{tabular}
	}
	\caption{\textbf{Design C sequences}. Each NS oligo is 48~nt long and lacks the FJ1 at the core. The NS sticky end is 8~nt long(\underline{CGATTGAC}). L oligos (40~nt) contain the complementary sticky end (\underline{GTCAATCG}) to the NS. Restriction enzyme sites are indicated: EcoRI (green) and EcoRV (magenta). Sequences with inverted RE sites between NS and L are also reported.}
	\label{tab:DesignC}
\end{table*}

\begin{table*}[h!]
	\centering
	\resizebox{0.8\textwidth}{!}{
		\begin{tabular}{|c|c|}
			\hline
			&Design D: NS(\EcoRI{\bf{EcoRI}}) + L(\EcoRV{\bf{EcoRV}}) \\
			\hline
			NS1 & $5^{\prime}-$CGTAAGGTCAATGCCCGCGGGCAATTCCTATT/\textbf{AA}/CGTAA\\
			& GGGGCATG\EcoRI{\bf{GAATTC}}CGCATCCTATTCC/\underline{CGATTGAC}$-3^{\prime}$\\
			NS2 & $5^{\prime}-$GGAATAGGATGCG\EcoRI{\bf{GAATTC}}CATGCCCCTTACG/\textbf{AA}/CCTAT\\
			&TCCCTTGG\EcoRI{\bf{GAATTC}}GATCCCGCGTAAG/\underline{CGATTGAC}$-3^{\prime}$\\
			NS3 & $5^{\prime}-$CTTACGCGGGATC\EcoRI{\bf{GAATTC}}CCAAGGGAATAGG/\textbf{AA}/AATAG\\
			&GAATTGCCCGCGGGCATTGACCTTACG/\underline{CGATTGAC}$-3^{\prime}$\\
			L1 & $5^{\prime}-$CGCATGA\EcoRV{\bf{GATATC}}CATTCAG/\underline{GTCAATCG}$-3^{\prime}$\\
			L2 & $5^{\prime}-$CTGAATG\EcoRV{\bf{GATATC}}TCATGCG/\underline{GTCAATCG}$-3^{\prime}$\\
			\hline
			&Design D: NS(\EcoRV{\bf{EcoRV}}) + L (\EcoRI{\bf{EcoRI}})  \\
			\hline
			NS1 & $5^{\prime}-$CGTAAGGTCAATGCCCGCGGGCAATTCCTATT/\textbf{AA}/CGTAA\\
			& GGGGCATG\EcoRV{\bf{GATATC}}CGCATCCTATTCC/\underline{CGATTGAC}$-3^{\prime}$\\
			NS2 & $5^{\prime}-$GGAATAGGATGCG\EcoRV{\bf{GATATC}}CATGCCCCTTACG/\textbf{AA}/CCTAT\\
			&TCCCTTGG\EcoRV{\bf{GATATC}}GATCCCGCGTAAG/\underline{CGATTGAC}$-3^{\prime}$\\
			NS3 & $5^{\prime}-$CTTACGCGGGATC\EcoRV{\bf{GATATC}}CCAAGGGAATAGG/\textbf{AA}/AATAG\\
			&GAATTGCCCGCGGGCATTGACCTTACG/\underline{CGATTGAC}$-3^{\prime}$\\
			L1 & $5^{\prime}-$CGCATGA\EcoRI{\bf{GAATTC}}CATTCAG/\underline{GTCAATCG}$-3^{\prime}$\\
			L2 & $5^{\prime}-$CTGAATG\EcoRI{\bf{GAATTC}}TCATGCG/\underline{GTCAATCG}$-3^{\prime}$\\
			\hline
		\end{tabular}
	}
	\caption{\textbf{Design D sequences}: Each NS oligo is 74~nt long and includes a FJ1 (\textbf{/AA/}). The NS sticky end is 8~nt long (\underline{CGATTGAC}). L oligos (28 nt long) contain a sticky end (\underline{GTCAATCG}), which is complementary to that of the NS. RE sites are indicated: EcoRI (green) and EcoRV (magenta). The oligo sequences with the RE sites inverted between NS and L are also reported.}
	\label{tab:DesignD}
\end{table*}

\clearpage

\begin{figure*}
	\centering
	\includegraphics[width=1\linewidth]{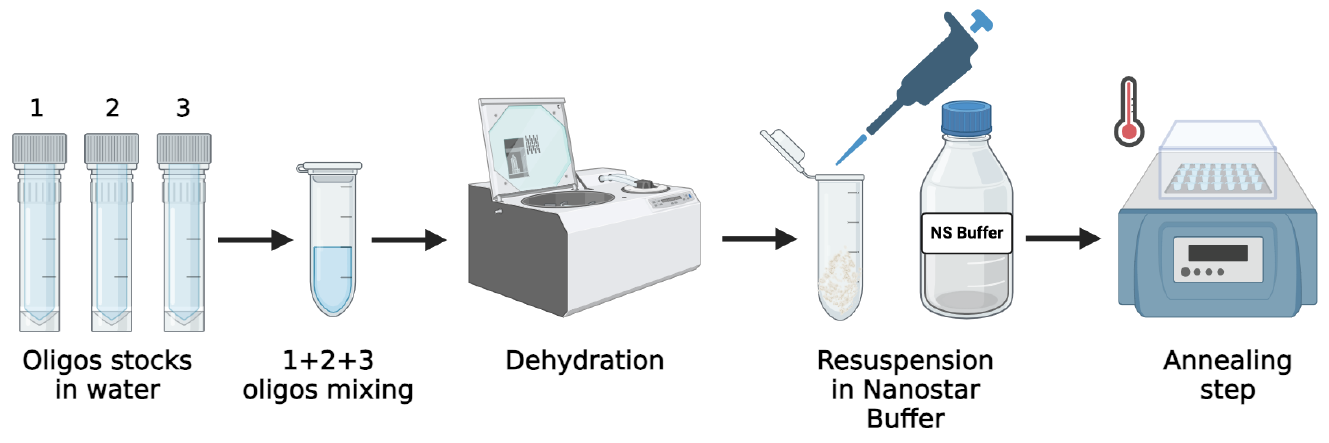} 
	\caption{\textbf{DNA Nanostar assembly}. Workflow for preparing nanostar samples (created with BioRender).} 
	\label{fig:NSprep}
\end{figure*}

\section{DNA Nanostars and Linkers assembly protocol}%
\label{sec:DNAnsassembly}
DNA oligonucleotides were purchased in dehydrated form from Integrated DNA Technologies (IDT). Following the provider's recommendations, tubes were centrifuged for 30~s to collect all material at the bottom. The oligos were rehydrated with Ultrapure Water to prepare 1~mM stock solutions (see Fig.~\ref{fig:NSprep}), which were typically stored in a refrigerator.

Equimolar amounts of the three oligos were combined from the stock solutions into a new Eppendorf\textsuperscript{\textregistered} DNA LoBind\textsuperscript{\textregistered} tube. The mixture was dried at 60\textdegree C for approximately 2~h in a vacuum concentrator (Eppendorf, Concentrator plus), with the tube open but covered by a sterile filter to prevent contamination. Once fully dehydrated, the sample was resuspended in Nanostar buffer (40~mM Tris, 40~mM sodium acetate, 1~mM EDTA, pH~8.0, and 150~mM NaCl)~\cite{Conrad2019}. The buffer volume added was adjusted to reach the desired final concentration.

To ensure complete dissolution of the sample, alternating cycles of heating at 60\textdegree C and vortexing were applied until no dry DNA was visible. The annealing process was then performed by heating the solution to 90\textdegree C ($T>T_{m1}$) for 2~min in a well-insulated heat block, followed by gradual cooling to 25\textdegree C ($T<T_{m3}$) over approximately 4~h. This slow cooling step was crucial for the correct assembly of the Y-shaped DNA structure. The resulting DNA nanostar solution was stored at 4\textdegree C until use.

Before experiments, the sample concentration was verified by measuring the absorbance at 260~nm (A$_{260}$) using a UV spectrophotometer (Thermo Scientific, NanoDrop Lite). Dilutions of 1:100 (A$_{260,1\%}$) and 1:1000 (A$_{260,0.1\%}$) were prepared in water, assuming that the absorbance of the 0.1\% dilution corresponded solely to ssDNA. Concentrations were calculated as described in references~\cite{Palombo2025,Conrad2019}.

For linker formation in designs~B-D, we prepared the stock solution in a separate tube from the DNAns. Equimolar amounts of the two ssDNA oligos were mixed in water, dehydrated, and resuspended in Nanostar buffer at a final concentration of 2~mM. The linker solution was annealed by cooling from 90 to 25\textdegree C to yield linear dsDNA. Its concentration was determined using a NanoDrop after 1:10 and 1:100 dilutions in Nanostar buffer. Mixing DNAns~B-D with their corresponding linker at a 1:1.5 molar ratio~\cite{Xing2011} produced a hydrogel through hybridization of complementary sticky ends.

\section{Sample Preparation for gel electrophoresis and microrheology}
\label{sec:samplePrep}
Standardized protocols were developed to ensure experimental consistency when investigating the behavior of the four DNAns designs, with and without restriction enzymes (REs), under various conditions. Three types of samples were prepared: (i) a Positive Control (PC) with low DNA concentration to verify the experimental setup; (ii) a Negative Control (NC) at both low and high DNA concentrations to confirm that observed effects originated from RE activity rather than external factors; and (iii) a Reaction Solution (RS) with high DNA concentration to evaluate RE efficiency in more concentrated or gel-like conditions.

\subsection{Positive Control}
\label{subsec:GELEP_PC}
Positive control samples were prepared following the guidelines in the “Optimizing Restriction Endonuclease Reactions” section of the NEB website~\cite{NEB}. This protocol was used for gel electrophoresis experiments.

Each 50~$\mu$L reaction contained a final DNA concentration of 0.4~$\mu$M, prepared by mixing in a single Eppendorf tube: (i) 38~$\mu$L of ultrapure water, (ii) 1~$\mu$L of DNA (DNAns or linker, at 1~$\mu$g/$\mu$L), (iii) 5~$\mu$L of 10$\times$ Nanostar Buffer to maintain constant ionic strength, (iv) 5~$\mu$L of 10$\times$ enzyme reaction buffer (rCutsmart for REs or DNaseI Reaction Buffer for DNaseI), and (v) 1~$\mu$L of the enzyme (replaced by the corresponding enzyme diluent buffer for the NC). The mixture was vortexed briefly and incubated in a heat block at 37~\textdegree C for 1~h. 

To stop digestion, 10~$\mu$L of 6$\times$ Purple Gel Loading Dye containing 60~mM EDTA was added to both PC and NC samples. The prepared solutions were then loaded into agarose gel wells for electrophoresis.

\subsection{Reaction Solution}
\label{sec:Reaction_Solution_procedure}
A different protocol was established for reaction samples prepared at higher DNAns concentrations, which was also applied to microrheology experiments.

For DNAns Design~A, samples were prepared at 250~$\mu$M DNAns (12.1~$\mu$g/$\mu$L) and incubated with different enzymes and units: 2~U of DNaseI, 20~U or 100~U of EcoRI-HF, 10~U of MspA1I, and 10~U of BtgI, in a final volume of 10~$\mu$L. These corresponded to enzyme-to-DNA mass ratios of 0.02~U/$\mu$g (DNaseI), 0.2~U/$\mu$g or 0.8~U/$\mu$g (EcoRI), and 0.08~U/$\mu$g (MspA1I and BtgI). The highest enzyme units available from NEB were used.
Each reaction was prepared by incubating an Eppendorf tube in a 60~\textdegree C heat block and sequentially adding: (i) 5~$\mu$L of DNAns at 500~$\mu$M (24.4~$\mu$g/$\mu$L) pre-heated at 60$^\circ$C for 2 minutes, (ii) 2.5~$\mu$L of 1$\times$ Nanostar Buffer, (iii) 1~$\mu$L of enzyme reaction buffer, (iv) 1~$\mu$L of enzyme (or diluent for NC), and (v) 0.5~$\mu$L of water (replaced by polystyrene beads in microrheology experiments). Due to the viscosity of the solutions, mixing was performed by pipetting 3–5 times using tips preheated to 60~\textdegree C to prevent premature cooling. The samples were incubated at 37~\textdegree C for 8–16~h, depending on enzyme stability. 
To stop digestion, 2~$\mu$L of 6$\times$ Purple Gel Loading Dye was added to each tube. Before opening, samples were briefly centrifuged to collect condensate at the bottom. Both NC and RS samples were then diluted in Nanostar Buffer to 25~ng/$\mu$L, and 10~$\mu$L of each (corresponding to 250~ng total DNA) was loaded per gel well.

For Designs~B-D, reactions were conducted at different enzyme-to-DNA ratios (0.45, 0.7, and 0.9~U/$\mu$g) by varying the total DNAns and linker mass while keeping enzyme units fixed at 100~U for EcoRI and EcoRV (maximum available from NEB). This allowed determination of the minimum enzyme-to-DNA ratio required for digestion at high DNA concentrations. Unlike Design~A, these systems included linkers that enabled lower DNAns concentrations while preserving viscoelastic properties~\cite{Xing2011}.
Two stock solutions were first prepared: 1~mM Nanostar (NS) and 2~mM Linker (L) in Nanostar Buffer. Both remained liquid at room temperature because their sticky ends were non-complementary. Gels (NS+L) were then formed by mixing NS and L in specific ratios (1:1.5~\cite{Xing2011}) to reach desired final concentrations within a total volume of 10~$\mu$L. For example, to achieve 0.45~U/$\mu$g, 3.3~$\mu$L of NS, 2.5~$\mu$L of L, and 4.2~$\mu$L of Nanostar Buffer were combined, yielding 0.33~mM NS and 0.50~mM L. For 0.68~U/$\mu$g, final concentrations were 0.23~mM (NS) and 0.35~mM (L), and for 0.9~U/$\mu$g, 0.16~mM (NS) and 0.24~mM (L). The NS+L gels, which were highly viscous, were heated at 60~\textdegree C (including tips) for 2~min.
Subsequently, 7.5~$\mu$L of the melted gel was mixed with (i) 1~$\mu$L enzyme reaction buffer, (ii) 1~$\mu$L enzyme (or diluent for NC), and (iii) 0.5~$\mu$L water (replaced by polystyrene beads in microrheology experiments). The samples were incubated at 37~\textdegree C for 21~h to optimize enzymatic digestion. To stop the reaction, 2~$\mu$L of 6$\times$ Purple Gel Loading Dye was added. The resulting samples were diluted in Nanostar Buffer to 25~ng/$\mu$L (if we consider only the mass of NS in the gel) or 45~ng/$\mu$L (considering the mass of both NS and L). Finally, 10~$\mu$L of each sample (250 or 450~ng total DNA) was loaded per gel well.

\section{Gel Electrophoresis}
\label{sec:gelelectrophoresis}
Gel electrophoresis is a widely used method for separating and analyzing charged DNA samples~\cite{Agarosegel}. The technique relies on measuring the migration of DNA molecules under an electric field, which correlates with their size and topology~\cite{Calladine1997,Bates2005}. Migration occurs through the pores of an agarose or polyacrylamide gel, whose pore size can be tuned by adjusting the polymer concentration. Smaller DNA fragments migrate faster and travel further than larger ones, as they experience less frictional resistance~\cite{Agarosegel}.

In this work, gel electrophoresis was performed using 3\% agarose gels to separate the small DNAns fragments. The gel matrix was prepared by dissolving UltraPure agarose powder in either TAE (Tris, acetic acid, EDTA) or LAB (lithium acetate borate) buffer, followed by heating until complete dissolution. The agarose solution was pre-stained with SYBR Safe dye (1:10,000 dilution) to enable visualization of DNA bands after electrophoresis. The molten solution was poured into a casting tray fitted with a comb to form wells and allowed to cool to room temperature.

Electrophoresis was carried out in a horizontal configuration at room temperature. After solidification, the gel was submerged in the same buffer used for its preparation (TAE or LAB) to ensure electrical conductivity and maintain stable pH. A voltage was applied across the gel, with the cathode at one end and the anode at the other. As DNA molecules are negatively charged they migrate toward the positive electrode. After electrophoresis, the separated DNA bands were visualized under a UV transilluminator and compared against a molecular weight marker. A low molecular weight DNA ladder (25–766~bp) was used, matching the expected sizes of the nanostars and linkers.

For experiments involving only DNAns (Design~A), the agarose gel was prepared by dissolving 3~g of agarose in 100~mL of 1$\times$ LAB buffer. Electrophoresis was performed at 150~V for 35~min, corresponding to an electric field strength of 7.5~V/cm (calculated as 150V/20cm, where 20~cm is the electrodes separation). The use of high voltage is enabled by lithium ions in the LAB buffer, which possess large hydration shells and low electrokinetic mobility, thereby reducing electrical conductance and minimizing heat generation during electrophoresis~\cite{Brody2004}.

For the other designs (B-D), a smaller gel tray was used, requiring 1.5~g of agarose in 50~mL of 1$\times$ TAE buffer. In this case, electrophoresis was conducted at 50~V for 70~min, corresponding to an electric field strength of 3.3~V/cm.

\begin{figure}[t]
	\centering
	\includegraphics[width=0.5\textwidth]{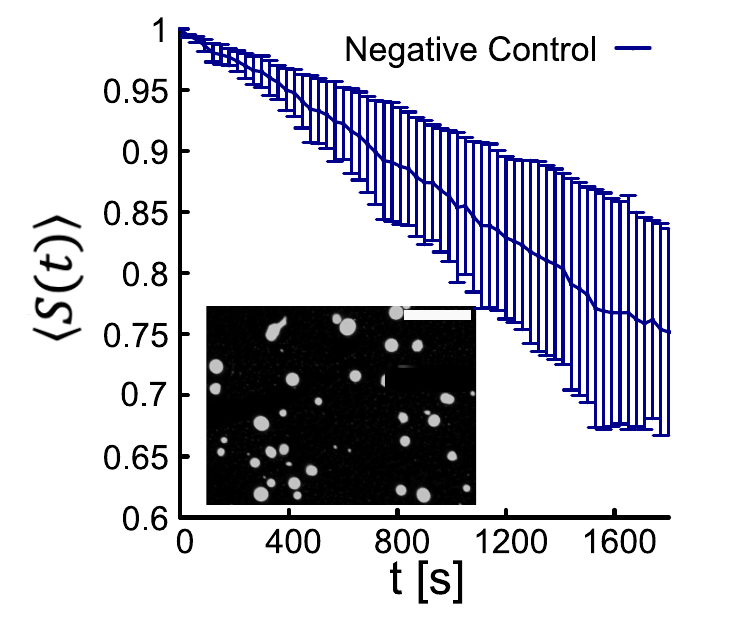}
	\caption{\textbf{Confocal imaging of DNAns droplets (Design~A).} YOYO-I–labelled droplets in a 2~$\mu$M solution before restriction enzyme addition (negative control; scale bar: 25~$\mu$m). The mean droplet size, $\langle S(t) \rangle$, decreases over time ($t$), indicating droplet shrinkage due to photobleaching.}
	\label{fig:Confocal_NC}
\end{figure}

\section{Confocal Microscopy}
\label{sec:confocalMicroscopy}
The spatial distribution and structural characteristics of the system at low DNAns concentrations were examined using confocal microscopy. This advanced fluorescence imaging technique provides high-resolution images and enables optical sectioning of samples. Unlike conventional widefield fluorescence microscopes that illuminate the entire specimen simultaneously, confocal systems focus a laser beam on a specific point within the sample and collect only the emitted light by excited fluorophores from that focal plane. A pinhole placed before the detector blocks out-of-focus light, significantly enhancing optical resolution (up to $\sim$200~nm laterally and $\sim$600~nm axially~\cite{Elliott2020}) and contrast. The size of the illuminated spot depends on the excitation wavelength, numerical aperture (NA) of the objective, and pinhole diameter.

In confocal microscopy, a laser scans a single optical plane by rastering the beam across the sample. By acquiring images at multiple focal depths (z-stacks), three-dimensional reconstructions of the sample can be generated. This capability is particularly valuable for studying thick or heterogeneous structures. Confocal systems can also image multiple fluorescent labels using distinct lasers and filters, allowing simultaneous visualization of different components within the same specimen.

All imaging was performed using a Zeiss LSM~700 Laser Scanning Microscope equipped with either a 40$\times$ oil-immersion objective (NA~1.3) or a 20$\times$ air objective (NA~0.4), depending on experimental requirements.

\subsection{Specimen preparation}
\label{sec:Specimen_preparation_confocal}
Our objective was to investigate the behavior of DNAns droplets formed at very low DNA concentrations, within the phase separation region~\cite{Saleh2020PNAS,Saleh2018SM}. We focused on Design~A nanostars to evaluate how the presence or absence of restriction enzymes influences droplet size, providing insight into the enzyme-responsive properties of the system. A stock solution was prepared by mixing 10~$\mu$L of DNAns (at 0.02~mM) with 1~$\mu$L of YOYO-I dye (1~mM) in a total volume of 100~$\mu$L. The mixture was gently rotated for 1~h to allow droplet formation while preventing sedimentation. Subsequently, 10~$\mu$L of the sample (with or without REs) was
prepared in a separate Eppendorf tube and then transferred to a sealed coverslip chamber and allowed to settle. Imaging was performed with a 40$\times$ oil-immersion objective (NA~1.3) using a 488~nm laser for YOYO-I excitation. All measurements were conducted at room temperature.

\begin{figure*}
	\centering
	\includegraphics[width=1\linewidth]{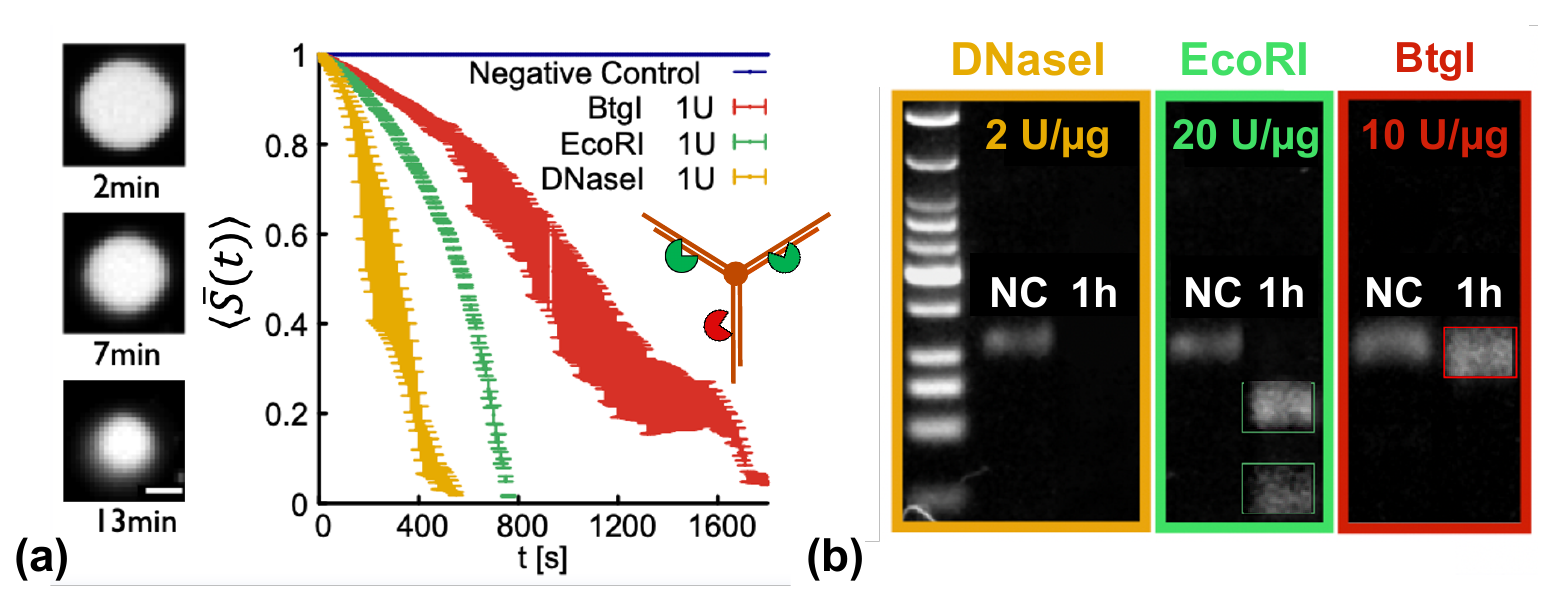}
	\caption{\textbf{RE-dependent DNAns cleavage within the phase boundary.} 
		\textbf{(a)} Left: Time-lapse confocal images showing the degradation of a YOYO-I–labeled droplet during EcoRI digestion (scale bar: 25~$\mu$m). Right: Normalized mean droplet size, $\langle \overline{S}(t) \rangle$, as a function of time. Degradation rates vary among enzymes: DNaseI (yellow) induces the fastest degradation, followed by EcoRI (green, two restriction sites) and BtgI (red, one restriction site), all resulting in complete digestion. 
		Inset: Schematic of the three-arm DNAns~A structure, showing a flexible core, flexible joints preceding 6-nt sticky ends, and 20-bp arms. 
		\textbf{(b)} Gel electrophoresis of DNAns solutions incubated for 1~h with DNaseI (2~U/$\mu$g, yellow), EcoRI (20~U/$\mu$g, green), and BtgI (10~U/$\mu$g, red).}
	\label{fig:Confocal_NC_enzymes}
\end{figure*}

\subsection{RE-dependent DNA Nanostar droplet degradation}
\label{subsec:ConfocalRE-dependent}
To directly visualize RE-mediated degradation of DNA nanostars, two complementary experiments were conducted: (i) time-resolved confocal imaging of DNAns droplets (2~$\mu$M) before and after RE addition, and (ii) gel electrophoresis of DNAns solutions (0.4~$\mu$M) incubated with and without REs (the latter replaced by enzyme diluent buffer in control samples). DNAns Design~A was used in both cases, with sample preparation protocols described in Sections~\ref{subsec:GELEP_PC} (positive control) and~\ref{sec:Specimen_preparation_confocal}.

Figure~\ref{fig:Confocal_NC} shows the evolution of the average droplet size, $\langle S(t) \rangle$, over time in the absence of restriction enzymes (negative control). Although no enzymatic activity is expected under these conditions, the droplets appear to shrink by approximately 30\% of their initial size after 30 minutes. This apparent decrease is attributed to photobleaching of the fluorescent signal during confocal imaging. As the fluorescence intensity gradually diminishes, the fixed-intensity threshold used in the image analysis algorithm underestimates droplet boundaries, leading to an apparent (but not physical) reduction in droplet size.

Consistent with previous findings~\cite{Saleh2020PNAS}, DNAns droplets shrank over time in the presence of REs due to DNA cleavage (Figure~\ref{fig:Confocal_NC_enzymes}\textbf{(a)}, left). To quantify this process while correcting for photobleaching (Figure~\ref{fig:Confocal_NC}), the average droplet size in the presence of REs, $\langle \overline{S}(t) \rangle$, was measured at various time points and normalized by the size in the negative control, $\langle S(t) \rangle$. As shown in Figure~\ref{fig:Confocal_NC_enzymes}\textbf{(a)} (right), $\langle \overline{S}(t) \rangle$ varied with enzyme type, revealing enzyme-dependent degradation kinetics: DNaseI (non-specific endonuclease) induced the fastest degradation, followed by EcoRI (two restriction sites per DNAns) and BtgI (one restriction site). All reactions used 1~U of enzyme per 0.38~$\mu$g of DNA, corresponding to an enzyme-to-DNA ratio of 2.6~U/$\mu$g.

To further confirm DNAns cleavage, the resulting fragments were analyzed by gel electrophoresis (Figure~\ref{fig:Confocal_NC_enzymes}\textbf{(b)}). In these experiments, 1~$\mu$g of DNAns was digested. The NC displayed a single band corresponding to intact nanostars, whereas RE-treated samples exhibited distinct digestion patterns: DNaseI (2~U/$\mu$g) fully degraded the DNA, leaving no visible bands; EcoRI (20~U/$\mu$g) generated two fragments of smaller size (highlighted in green); and BtgI (10~U/$\mu$g) produced a single shifted fragment (highlighted in red).

These results confirm that DNAns within the phase-separation region undergo enzyme-specific degradation characterized by distinct kinetics and cleavage patterns depending on the restriction enzyme used.

\section{Effect of glycerol on the rheology of DNA nanostars}
\label{sec:glyceroleffect}
\begin{figure}[t!]
	\centering
	\includegraphics[width=1.0\linewidth]{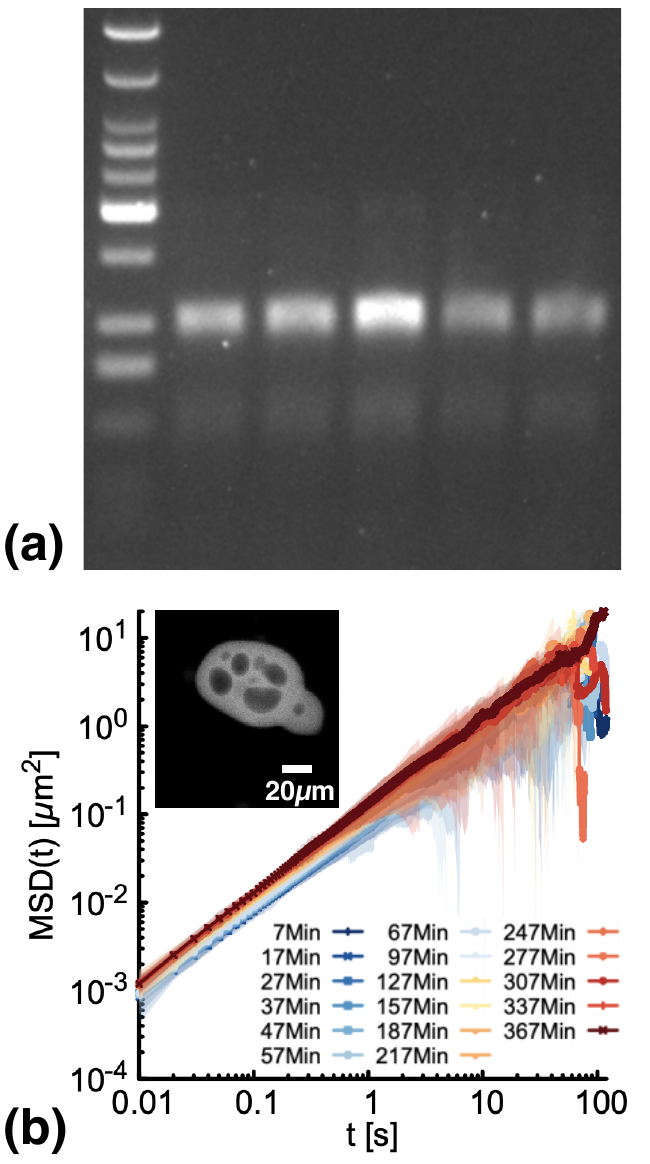}
	\caption{\textbf{Negative control for Design~A samples.} 
		\textbf{(a)} Gel electrophoresis of the negative control (no enzyme added) for DNAns gels (Design~A, 250~$\mu$M). No degradation is observed under DNaseI storage buffer after incubation at 35~\textdegree C for 0, 1, 2, 4, and 6~h. 
		\textbf{(b)} Time-resolved microrheology of the same sample. The MSD$(t)$ exhibits a predominantly liquid-like response with a slight upward shift over time (see also Figure~\ref{M-fig:MSD_RE}\textbf{(c)}). Inset: Confocal image showing the effect of glycerol on DNAns, with pore formation visible in droplets at 2~$\mu$M (scale bar: 20~$\mu$m).
	}
	\label{fig:glycerol}
\end{figure}

To establish a baseline, the system was first characterized in the absence of enzyme (negative control). This ensured that any fragments or rheological changes observed in our experiments could be attributed solely to enzymatic activity. For gel electrophoresis, a master solution of Design~A DNAns (250~$\mu$M, $\sim$120~$\mu$g/$\mu$L) was prepared, in which the enzyme was replaced with its corresponding diluent buffer (1~mM Tris, 0.2~mM CaCl\textsubscript{2}, 5\% glycerol, pH~7.6 at 25~\textdegree C). The solution was incubated at 35~\textdegree C and aliquots were taken at specific time points. After six hours, a single band remained visible on the gel (Figure~\ref{fig:glycerol}\textbf{(a)}), confirming that the nanostars remained intact.

Microrheology measurements supported this finding (Figure~\ref{fig:glycerol}\textbf{(b)}). Under identical conditions, 200~nm tracer particles were tracked every 10~min for the first hour and every 30~min thereafter, up to six hours. Surprisingly, the samples exhibited predominantly liquid-like behavior, rather than the viscoelastic response reported previously~\cite{Palombo2025}. This behavior is likely due to the presence of 5\% glycerol in the buffer, which may displace water molecules and weaken hydrogen bonds between nanostars~\cite{Glycerol}. Confocal imaging of 2~$\mu$M DNAns droplets corroborated this hypothesis, showing pore formation upon glycerol addition (Figure~\ref{fig:glycerol}(b), inset). In addition, a slight upward shift in the mean squared displacement over time was observed, suggesting that glycerol modifies the network structure. The corresponding viscosity, derived from the MSD analysis, is shown in Figure~\ref{M-fig:MSD_RE}\textbf{(c)} of the main text.

\section{Non-specific enzyme DNase~I digests all DNAns gel designs}
\label{sec:dnase1DiffDesigns}
\begin{figure*}
	\centering
	\includegraphics[width=1.0\textwidth]{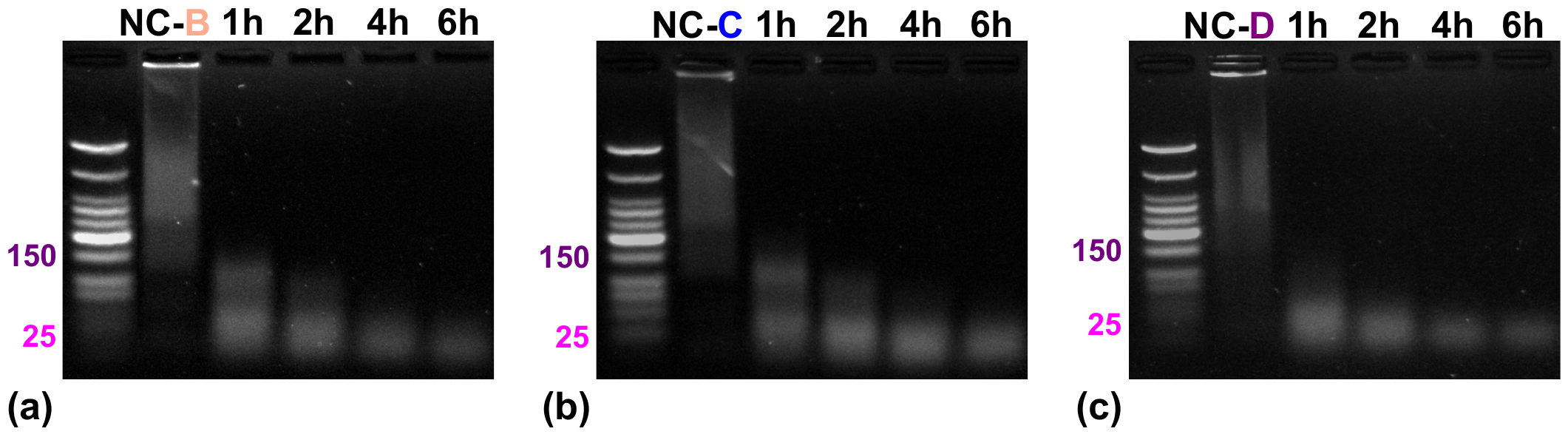}
	\caption{\textbf{DNase~I degradation of different DNAns designs.} 
		Gel electrophoresis showing the degradation of DNAns hydrogels by DNase~I over time (at 0.02~U/$\mu$g and 35~\textdegree C). Gel images display DNA digestion at 1, 2, 4, and 6~h. The 150~bp violet band corresponds to intact DNAns, while the 25~bp magenta band indicates smaller fragments resulting from enzymatic degradation. Data are shown for Designs~B~\textbf{(a)}, C~\textbf{(b)}, and D~\textbf{(c)}.
	}
	\label{fig:DNaseIdiffDesigns}
\end{figure*}

In the main text, we demonstrated that DNase~I efficiently digests gels composed of Design~A DNAns. Here, we confirm that DNase~I exhibits comparable efficiency in digesting DNAns solutions across the other three designs (B–D), as illustrated in Figure~\ref{fig:DNaseIdiffDesigns}. For all three designs, digestion was performed at 35~\textdegree C over four time points (1, 2, 4, and 6~h) using the same enzyme-to-DNA ratio of 0.02~U/$\mu$g, where $\mu$g refers to the total DNA mass of nanostars plus linker.

As shown in Figure~\ref{fig:DNaseIdiffDesigns}, the DNA in the negative controls for Designs~B–D remains trapped in the well, likely due to the presence of longer sticky ends (8~nt versus 6~nt in Design~A) and the linker, which promote the formation of more stable and larger structures even at low concentrations~\cite{Sato2020,Franco2022}. Upon DNase~I addition, smaller DNA fragments appear in the gel, indicating complete digestion of the samples.

\section{ELISA}
\label{sec:elisa}
Based on the estimated pore size of the design D hydrogel, immunoglobulin G (IgG) antibodies were identified as compatible for capture and retention in DNAns hydrogels. IgG antibodies have approximate dimensions of 14.5~nm $\times$ 8.5~nm $\times$ 4.0~nm~\cite{Yih2008}. The antibody used in this study was Bevacizumab (Bio-Rad, MCA6089), a humanized monoclonal IgG that inhibits vascular endothelial growth factor (VEGF) and is used clinically for cancer therapy.

\paragraph{Hydrogel preparation.} 
Two stock solutions were prepared: one containing only design D DNAns and one containing only linker (28~nt). Antibody was gently mixed with the linker solution at room temperature to achieve final concentrations of 1.68~mM linker and 0.114~$\mu$g/$\mu$L antibody. This mixture was then combined with the DNAns solution to form hydrogels with final concentrations of 0.33~mM DNAns, 0.5~mM linker, and 0.034~$\mu$g/$\mu$L antibody. To facilitate handling, the DNAns solution was heated to 45~$^\circ$C for 5~minutes prior to mixing. Aliquots of 30$\mu$L hydrogel were placed at the bottom of 1.5~mL microfuge tubes. Negative control hydrogels (10~$\mu$L) were prepared using the same method but without antibody. Nine replicates of each hydrogel variant were prepared.

\paragraph{Hydrogel washing.}
Hydrogels were washed three times to remove unbound antibody and excess DNA. For each wash, three hydrogel volumes of nanostar buffer were gently added and removed without disturbing the hydrogel. Duplicate samples of the spent buffer from each wash were collected and stored at $-20^\circ$C for later analysis.

\begin{figure}[b!]
\centering
\includegraphics[width=1.0\linewidth]{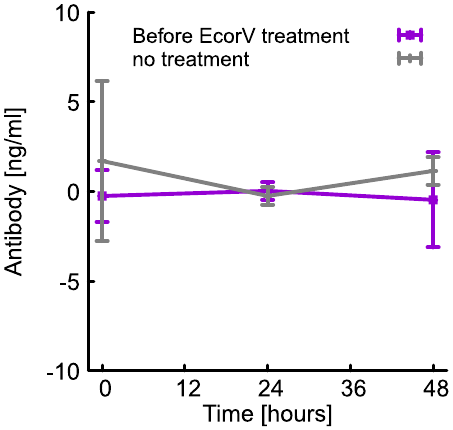}
\caption{\textbf{Retention of IgG antibodies in design D DNA hydrogels in the absence of enzymatic activity.} Antibody release profiles measured by ELISA for untreated hydrogels.}
\label{fig:elisaNC}
\end{figure}

\paragraph{Antibody retention assays.}
To assess retention, nineteen hydrogel volumes of nanostar buffer were added to each hydrogel and incubated at room temperature. At 0, 24, and 48~hours, duplicate samples of buffer were collected. For each collection, the buffer was carefully transferred to a fresh tube, vortexed, sampled in duplicate, and returned to the top of the hydrogel. Each sample volume matched the hydrogel volume and was stored at $-20^\circ$C until analysis. Results from ELISA experiments in Fig.~\ref{fig:elisaNC} show that on average antibody levels never exceeded 2~ng/ml within the first 48 hours.

\paragraph{Controlled release assays.}
For controlled release experiments, hydrogels with and without antibody were treated with EcoRV-HF or no enzyme, all in triplicates. Prior to enzyme treatment, hydrogels were washed three times as described above. For treatment, nine hydrogel volumes of nuclease mixture were added to each hydrogel, and samples were incubated at room temperature. EcoRV-HF samples contained 0.68~U EcoRV-HF in 1$\times$ rCutSmart buffer, and no-enzyme controls were treated with water. Over a 73-hour time course, duplicate samples of equal volume to the hydrogel were collected as described above. Each sample was treated with 0.1~M EDTA to achieve a final concentration of 0.01~M for nuclease inactivation and stored at $-20^\circ$C until analysis.

\paragraph{DNA integrity analysis.}
Enzymatic degradation of hydrogels was analyzed by agarose gel electrophoresis. A 3\% agarose gel was prepared in TAE buffer, and 1~$\mu$L of samples collected during the nuclease treatment time course was loaded. Gels were run at 50~V for 90~minutes, corresponding to a field strength of 3.1~V/cm. Following electrophoresis, gels were post-stained with 1$\times$ SYBR Safe in 150~mL TAE and imaged using a Bio-Rad Molecular Imager\textsuperscript{\textregistered} Gel Doc\textsuperscript{TM} XR+ Imaging System with Image Lab software v6.1.

To directly assess hydrogel integrity, the intensity of the lowest–molecular-weight band, corresponding to DNA fragments released into solution, was quantified using ImageJ/Fiji. Prior to analysis, gel images were despeckled and background-subtracted to ensure uniform noise across both axes. Band intensities were manually selected using local intensity minima surrounding each band, and background signal was estimated from five randomly chosen regions lacking detectable bands and subtracted from each measurement. To enable quantitative comparison across different gels, the intensity of the lowest band in each sample lane was normalized by the intensity of the lowest band in the corresponding DNA ladder, which provides a consistent reference concentration across gels.
\bibliography{bibliography}